\documentclass[12pt]{iopart}

\usepackage{iopams}  

\usepackage{graphicx}
\usepackage{braket}
\usepackage{array}
\usepackage{qcircuit}
\usepackage[ruled,vlined]{algorithm2e}
\usepackage[table]{xcolor} 
\newcommand{\kk}{\mathbf{k}}
\newcommand{\pp}{\mathbf{p}}
\usepackage{makecell}
\usepackage[table]{xcolor}

\begin{document}
\newcommand{\question}[1]{\textcolor{blue}{#1}}

\title{Simulating Quantum Materials with Digital Quantum Computers}

\author{Lindsay Bassman, Miroslav Urbanek, Mekena Metcalf, Jonathan Carter}
\address{Lawrence Berkeley National Laboratory, Berkeley, CA 94720, USA}
\ead{lbassman@lbl.gov}
\author{Alexander F. Kemper}
\address{Department of Physics, North Carolina State University, Raleigh, North Carolina 27695, USA}
\author{Wibe de Jong}
\address{Lawrence Berkeley National Laboratory, Berkeley, CA 94720, USA}

\vspace{10pt}

\begin{abstract}
Quantum materials exhibit a wide array of exotic phenomena and practically useful properties.  A better understanding of these materials can provide deeper insights into fundamental physics in the quantum realm as well as advance technology for entertainment, healthcare, and sustainability.  The emergence of digital quantum computers (DQCs), which can efficiently perform quantum simulations that are otherwise intractable on classical computers, provides a promising path forward for testing and analyzing the remarkable, and often counter-intuitive, behavior of quantum materials.  Equipped with these new tools, scientists from diverse domains are racing towards achieving physical quantum advantage (i.e., using a quantum computer to learn new physics with a computation that cannot feasibly be run on any classical computer).  The aim of this review, therefore, is to provide a summary of progress made towards this goal that is accessible to scientists across the physical sciences.  We will first review the available technology and algorithms, and detail the myriad ways to represent materials on quantum computers. Next, we will showcase the simulations that have been successfully performed on currently available DQCs, emphasizing the variety of properties, both static and dynamic, that can be studied with this nascent technology.  Finally, we work through two examples of how to map a materials problem onto a DQC, with full code included in the Supplementary Material.  It is our hope that this review can serve as an organized overview of progress in the field for domain experts and an accessible introduction to scientists in related fields interested in beginning to perform their own simulations of quantum materials on DQCs.
\end{abstract}
\maketitle
\tableofcontents

\section{Introduction}
Quantum materials \cite{keimer2017physics, giustino20202020}, such as superconductors and complex magnetic and topological materials, exhibit properties and behaviors which can only be described by the laws of quantum mechanics.  A better understanding of the sometimes counter-intuitive characteristics of these materials holds great promise for revolutionizing technology for information processing tasks such as computation, sensing, communication, and metrology, as well as efficient energy storage and pharmaceutical drug development \cite{basov2017towards,tokura2017emergent, degen2017quantum, han2018quantum, cao2018potential}.  Furthermore, advancing our knowledge of quantum materials at a fundamental level can have far-reaching and unforeseeable consequences in terms of better understanding the laws of nature as a whole.  

Because accurate measurement of intrinsic properties and response mechanisms of quantum materials can be difficult to achieve with experiment, computer simulation is often utilized in a predictive capacity to obtain approximations of and insights into various features of quantum materials \cite{georgescu2014quantum, head2020quantum}.  Over the past half century, such simulations performed on classical computers have been instrumental in advancing such fields as quantum chemistry, materials science, and condensed matter physics \cite{freeman2002materials, steinhauser2009review, schleife2014quantum}.  Simulations of quantum materials require representing all or parts of the system by a wavefunction, as opposed to classical positions and momenta. There is, however, an inherent difficulty with wavefunction-based materials simulations on classical computers, namely, the complexity of simulating a many-body wavefunction grows exponentially with system size.  The complexity of a simulation refers to the amounts of compute time and/or memory resources that are required.  When the size of these resources grows exponentially with the size the system (e.g. number of electrons, number of basis states) the simulation is considered to be inefficient (whereas polynomial scaling is considered efficient).  The inefficient scaling of exact quantum simulations on classical computers causes systems with only several atoms to quickly become intractable \cite{kohn1999nobel}.  

Many approximate simulation methods, notably density functional theory \cite{hohenberg1964inhomogeneous, kohn1965self, jones2015density}, have been developed to reduce the simulation complexity to polynomial (i.e., efficient) scaling with system size.  Reductions in computational complexity generally stem from making approximations that eliminate the need to store or manipulate an explicit representation of the exact many-body wavefunction of the material system.  Such procedures allow for reasonably accurate simulation results for certain classes of materials with system sizes reaching hundreds to thousands of atoms \cite{shimamura2014hydrogen, lin2017ultrafast}.  Unfortunately, these methods cannot be extended to all quantum materials simulations, either due to high accuracy requirements, high degrees of entanglement within the system, or large system sizes.  

One potential path forward is to perform quantum materials simulations on quantum computers.  The theoretical inception of simulating quantum systems on quantum computers dates back to a lecture given by Richard Feynman in the early 1980s \cite{feynman1982simulating}. The most widely used quantum computers store information in two-level quantum systems called quantum bits, or qubits.  In principle, more levels are accessible, and information can be encoded in a three-level system (qutrit) or higher \cite{nielsen2010quantum}, or as a continuous variable in optical quantum systems \cite{neergaard2010optical}.  Due to their quantum nature, which enables superposition and entanglement, qubits can be programmed to efficiently store the wavefunction of a quantum system.  Furthermore, evolution of the system can be efficiently simulated as long as the system Hamiltonian is only comprised of local interactions \cite{lloyd1996universal}.  Fortunately, this is not too stringent of a requirement for materials simulations as ``any system that is consistent with special and general relativity evolves according to local interactions'' \cite{lloyd1996universal}.  As the primary intention of materials simulations on quantum computers is to better understand real materials for human use, we may presume all Hamiltonians of interest will be local, and thus efficient to simulate on a quantum computer.

There are a number of candidate implementations for quantum computers, many of which are under active investigation.  Currently available quantum computers, referred to as noisy intermediate-scale quantum (NISQ) computers \cite{preskill2018quantum}, are limited in their total number of qubits, as well as in the fidelity of information stored in and operations performed on those qubits.  While complex error-correcting schemes may one day give rise to fault-tolerant quantum computers, quantum computers of the near future will have too few qubits for such overhead.  Given these constraints, highly anticipated uses for quantum computers like factorizing large numbers \cite{shor1999polynomial} and searching unsorted databases \cite{grover1996fast} are not currently viable, making materials simulations one of the most promising applications for NISQ computers.

A great deal of theoretical progress has been made in algorithm development for materials simulations on quantum computers \cite{cao2019quantum, bauer2020quantum, mcardle2020quantum, head2020quantum}.  Since the foundational algorithm for simulating many-body Fermi systems on quantum computers was published over twenty years ago \cite{abrams1997simulation}, myriad algorithms have been proposed for simulating static properties, like ground-state energy, as well as for simulating quantum dynamics.  The last five years has seen incremental improvements in complexity bounds and resource requirements of these algorithms.  

Implementing such algorithms on quantum computers requires mapping components of the algorithms onto the qubits of the quantum computer.  There are numerous ways to encode a material system into qubits, with the optimal one determined by the specifics of the simulated material \cite{kassal2011simulating}.  A variety of software packages have been developed to aid in implementing such algorithms and encodings, which are slowly growing and maturing \cite{fingerhuth2018open, heim2020quantum}.  However, due to the limitations of current NISQ hardware, successful simulations of real materials on quantum computers have been limited to proof-of-concept simulations of small molecules and toy material models.  As more efficient algorithms reduce requirements of qubit number and fidelity, and better quantum computing hardware increases numbers of higher quality qubits, it is believed that quantum materials simulations on quantum computers of the near-future can aid in discovering new materials via high-throughput simulations, elucidating material behaviors and mechanisms, rationalizing experimental results, and exploring new physical models and materials theories.

Simulations of quantum materials break into two main paradigms: (i) \emph{static} simulations that estimate intrinsic quantum material properties such as ground and excited state energies and (ii) \emph{dynamic} simulations that observe how time-dependent properties change as the quantum material evolves in time.  Static simulations are employed for problems in quantum chemistry and quantum molecular spectroscopy, as well as computing correlated electronic structure in materials.  Quantum chemistry focuses on determining the lowest-lying energy states of molecules.  Such information can shed light on the electronic and structural characteristics of enzyme active sites \cite{kurashige2013entangled, sharma2014low, cao2018protonation}, intermediate geometries and molecular adsorption energies in catalytic reactions \cite{norskov2009towards,schimka2010accurate, wodtke2016electronically}, and the photo-chemistry in light-harvesting processes \cite{segatta2019quantum}.  

Whereas quantum chemistry problems make the key approximation of freezing the positions of the nuclei in the molecule, known as the Born--Oppenheimer approximation, quantum molecular spectroscopy must consider the atomic motion of the nuclei.  Specifically, it requires constructing a Hamiltonian to quantum mechanically describe the motion of the nuclei, which is difficult to compute due to the Hamiltonian's dependence on the potential energy surfaces generated by the electrons \cite{christiansen2007vibrational, christiansen2012selected}.  Once this nuclear Hamiltonian is formulated, it can be used to compute the the (ro-)vibrational spectra of molecules which can elucidate molecular structure and behavior \cite{csaszar2012fourth}.

Moving from molecules to larger crystalline materials opens the door to a more diverse set of properties and systems that can be simulated.  Systems like Mott insulators \cite{raghu2008topological}, high-temperature superconductors \cite{dagotto1994correlated}, two-dimensional materials \cite{bhimanapati2015recent}, and frustrated spin systems \cite{diep2013frustrated} all exhibit interesting quantum properties intimately related to the high level of correlation amongst their comprising electrons.  The fields of condensed matter physics and materials science are largely interested in computing electronic band structures and phase diagrams to better understand these strongly correlated materials.

Dynamic simulations, on the other hand, are essential for the study of time-dependent properties in materials.   Specifically, these simulations are concerned with electronic and nuclear motion through time to discern rates of chemical processes \cite{suleimanov2016chemical}, elucidate electron transport \cite{mallajosyula2008sequence}, and unravel the dynamical interactions of matter with light \cite{bassman2018electronic}.  Two major areas of interest for such simulations include (i) non-adiabatic quantum effects, such as non-radiative energy relaxation processes \cite{lin2017ultrafast}, and (ii) non-equilibrium dynamics for explaining equilibration of quantum systems \cite{dawson2018equilibration} and describing exotic driven and topological materials \cite{tokura2017emergent, oka2019floquet}.

This review is organized as follows.  In Section \ref{sec:technology} we give an overview of the technology available for simulating materials on quantum computers, including both hardware and software.  In Section \ref{sec:algorithms}, we provide an overview of the algorithms that have been developed and successively improved upon for such simulations.  In Section \ref{sec:qubit_rep}, we summarize the various Hamiltonians that have been used to model materials and how they can be mapped onto the qubits of the quantum computer.  In Section \ref{sec:static}, we describe the static materials simulations that have been successfully carried out on quantum computers, while Section \ref{sec:dynamic} covers dynamic simulations.  In Section \ref{sec:examples} we work through full examples for performing a sample static and dynamic material simulation on a quantum computer.  Finally, in Section \ref{sec:conclusion} we conclude with an outlook on future progress on all fronts for simulation of materials on quantum computers.  

\section{Available Technology}
\label{sec:technology}
Over the last decade, a tremendous amount of technological progress has been made in both the hardware and software that are available for performing materials simulations on quantum computers.  The quantum hardware we focus on in this review comes in the form of digital quantum computers (as opposed to adiabatic or analog quantum computers \cite{das2008colloquium, georgescu2014quantum, gross2017quantum, albash2018adiabatic, hauke2020perspectives}), comprising a set of qubits on which a universal set of quantum logic gates can be enacted.  We therefore highlight software libraries and full-stack software packages for classical computers that have been developed for designing, optimizing, and executing quantum circuits for execution on digital quantum computers.  A steady stream of advances unfolding on both fronts feed into the recent progress that has been seen in materials simulations on quantum computers.  Hardware improvements, including increasing the total number of qubits, enhancing qubit connectivity, and raising gate fidelity, allow for larger quantum circuits, and hence more complex materials, to be simulated.  Improvements in software, particularly those which enable shorter circuits, allow for higher-fidelity results from the currently available quantum hardware.  In this section, we provide an overview of the state-of-the-art hardware and software for simulating materials on quantum computers.

\subsection{Hardware}
In principle, any quantum system can be used for quantum information processing~\cite{Monroe:2002, Nakahara:2008, nielsen2010quantum, Buluta:2011}.  However, what distinguishes a good candidate quantum computer is (i) long coherence times, (ii) fast gate operation speeds, and (iii) high gate fidelities.  The coherence time of a quantum computer is given by the length of time its qubit states can be stored faithfully.  In order to achieve reliable results, a quantum computer should complete a given quantum circuit in a wall-clock time shorter than the coherence time of the qubits.  Quantum circuits that run longer than the coherence time lead to error-prone results.  Therefore, fast gate operations are desirable, enabling more gates to be performed within the coherence time.  The gate fidelity refers to the probability that the gate performs its operation error-free.  As in classical computing, logic gates come with associated error rates, and the goal is to make these rates as low as possible.  This is of particular importance in the NISQ era, where qubit numbers are too low to afford advanced error-correcting techniques.

Currently, there are several contending technologies for implementing qubits for digital quantum computers.  The most widely available quantum computers, which are accessible over the cloud, include those based on superconducting qubits (IBM \cite{QC_IBM}, Rigetti \cite{QC_Rigetti}), trapped-ion qubits (IonQ \cite{QC_IonQ}, Honeywell \cite{QC_Honeywell}), and photonic qubits (Xanadu \cite{QC_Xanadu}).  All have been scaled to tens of qubits, with plans to scale to a thousand qubits within three years \cite{gambetta2020ibm}.  These, along with emerging qubit technologies are described in the sections below.

\subsubsection{Superconducting Qubits}
Superconducting qubits represent one of the most promising quantum hardware technologies, with many such devices under current development in academic, government, and industry settings \cite{krantz2019quantum}.  Such qubits are comprised of superconducting solid-state circuits, engineered to provide good coherence times and control abilities ~\cite{Schoelkopf:2008, Clarke:2008, Ladd:2010, Siddiqi:2011, You:2011, Barends:2014, Wendin:2017}. Their main advantage is that as solid-state systems they are easy to control with electronic devices. However, to achieve superconductivity and to decrease noise to acceptable levels, they must be cooled to milli-Kelvin temperatures.

Superconducting qubits typically consist of a capacitor, an inductor, and a Josephson junction. The capacitor and inductor create an LC oscillator. This superconducting oscillator is a quantum harmonic oscillator with discrete energy levels. A quantum harmonic oscillator has equidistant energy levels. It is therefore difficult to induce a transition between a particular pair of states. By adding a Josephson junction, which is a nonlinear inductor, the energy levels are shifted into non-equidistant levels. Only then it is possible to uniquely address a transition between the ground state and the first excited state. Similar to other physical realizations, superconducting qubits have higher excited states, which can be ignored or used as multilevel quantum registers called qudits~\cite{Neeley:2009, Bianchetti:2010}.

Circuits with different properties can be created by changing the capacitance and inductance of the circuit components. Examples include phase qubits~\cite{Martinis:2002, Steffen:2006}, flux qubits~\cite{Mooij:1999}, charge qubits~\cite{Nakamura:1999}, transmons~\cite{Koch:2007, Paik:2011}, Xmons~\cite{Barends:2013}, and gatemons~\cite{Casparis:2016}. Qubits are typically coupled to electromagnetic cavities and their control and readout is performed by microwave pulses into these cavities~\cite{Roch:2012, Johnson:2012, Riste:2012, Vijay:2012, Murch:2013, Riste:2013, Roch:2014, Jeffrey:2014, Sun:2014, Weber:2014, Eichler:2014, OBrien:2014, Sank:2016, Whaley:2015}. There has been rapid improvement in this technology in the past decade~\cite{Vijay:2009, Devoret:2013}.  There are currently a handful of functional quantum computers based on superconducting qubits, with some containing up to 72 functional qubits.  Superconducting qubits typically have coherence times around 50--100 $\mu$s, though have demonstrated coherence times of up to 8~ms \cite{Earnest:2018}.  Gate speeds for superconducting qubits are on the order of tens of ns, while two-qubit gate fidelities up to 99.5\% have been achieved \cite{kjaergaard2020superconducting}.

\subsubsection{Trapped-Ion Qubits}
Another extremely promising technology for digital quantum computers is based on trapped-ion qubits  \cite{Brown:2016, Schafer:2018,bruzewicz2019trapped}.  Already, several trapped-ion based quantum computers, with up to 32 qubits, have been implemented with limited access via the cloud. In a trapped ion-quantum computer, ions confined in lattice traps formed by electromagnetic fields serve as the qubits.  The internal electronic states of the ions are mapped to the qubit states, while shared motional modes of ions allow for transfer of quantum
information between ions to enable qubit entanglement.  Trapped ions have the highest coherence times among all technological candidates, with coherence times longer than 10 minutes observed \cite{Wang:2017}.  Another advantage of trapped-ions is the high-fidelity of both one- and two-qubit gates. Single-qubit rotations have been shown to obtain fidelities of over 99.99\% \cite{harty2014high, Ballance:2016}, significantly above the threshold level believed to be required for fault-tolerant quantum computation. Furthermore, fidelities of up to 99.9\% have been experimentally demonstrated for two-qubit entangling gates \cite{Ballance:2016, gaebler2016high}.  

The disadvantage of trapped-ions is that gate speeds are significantly slower; while execution times for two-qubit gates on trapped-ions have been shown to be as fast as 1.6 $\mu$s \cite{Schafer:2018}, analogous gates on superconducting qubits can be executed in times two orders of magnitude shorter. Furthermore, scalability is also a concern as the control complexity grows with the square of number of ions~\cite{Popkin:2016}. One solution may be to use them in a modular architecture, with each module containing only a limited number of qubits~\cite{Hucul:2014}.
 
\subsubsection{Photonic Qubits}
Optical photons are another suitable realization of qubits~\cite{Obrien:2007, OBrien:2009, Shadbolt:2011, AspuruGuzik:2012, Flamini:2019}. The quantum information can be encoded in photon path, polarization, spatial modes or in time~\cite{Flamini:2019}. The main obstacle with photonic qubits is the limited interaction between photons.  Nonlinear optical media are typically used to facilitate interactions between photons. Nonlinear interaction, however, is weak and can also lead to absorption. There exists another approach using only linear elements with help from measurement of ancilla qubits \cite{Knill:2001, Kok:2007, Carolan:2015}. Universal quantum computation is then possible with single-photon generators, phase shifters, beam splitters, and photon detectors. However, the required resources are high.  As such, photonic quantum technologies are utilized in quantum communication, i.e., to transfer quantum information over long distances.

\subsubsection{Emerging Technologies}
Whereas the previously mentioned qubit technologies have seen successful implementations that are accessible over the cloud for scientific and commercial use, other emerging technologies are under active development.  The next qubit technology closest to commercial release is based on cold, neutral atoms \cite{deutsch2000quantum, negretti2011quantum, saffman2016quantum, briegel2000quantum, henriet2020quantum}.  Here, laser-cooled, neutral atoms are arranged in a lattice using optical or magnetic traps.  Qubits states are encoded in either the Zeeman or hyperfine ground states, which provide long coherence times.  One-qubit gates can be executed with microwaves or two-frequency Raman light, while two-qubit entangling gates can be implemented via Rydberg interactions or controlled collisions.   Long coherence times are a major advantage of neutral atom qubits, with coherence times up to an hour having been observed in cryogenic environments \cite{willems1995creating}.  Another advantage of neutral atom qubits is the ability to natively implement multi-qubit gates, which can greatly improve quantum circuit depth and error tolerance.  Two disadvantages of neutral atoms are slow gate speeds, which are on the order of $\mu$s, and low two-qubit gate fidelities, around 94.1\% \cite{levine2018high}.

Another emerging technology is the use of nitrogen-vacancy centers~\cite{Neumann:2008, Balasubramanian:2009, Weber:2010, Doherty:2013} as qubits.  These are created by replacing a portion of carbon atoms with nitrogen atoms in the crystal lattice of the diamond. The quantum information is stored in the nuclear spin. The advantages of such nuclear spins are their long coherence times and the fact that they can be stored at room temperature.  A final promising technology is topological qubits~\cite{Freedman:2001, Kitaev:2003}, which, by design, are insensitive to environmental noise, one of the largest sources of decoherence in qubits. Topological qubits can be realized by quasi-particles that are neither bosons or fermions but obey anyonic statistics. An example of such quasiparticles are Majorana zero modes~\cite{Sau:2017}, which have been physically realized in Majorana nanowires~\cite{Albrecht:2016, Gul:2018}. Scalable designs of topological quantum computers have been proposed~\cite{Karzig:2017}, however, the underlying technology is not yet ready for practical quantum applications.

\subsection{Software}
As new implementations of quantum hardware have evolved to have larger qubit counts with greater gate fidelities, there has been increased activity in the development of software to support quantum program development in the form of programming libraries and full software stacks.  Specifically, software is needed to design the quantum circuit that carries out the materials simulation, such as initial state preparation, appropriate system evolution, and measurement of the desired observable.  Furthermore, if the circuit is designed to run on a realistic backend with fixed qubit connectivity and noisy qubits and gates, classical software is also needed for quantum circuit optimization, which enforces the qubit topology in the circuit and attempts to minimize the circuit depth.  This quantum circuit optimization step is referred to as quantum compilation or circuit synthesis.  Recently, an in-depth review outlined the functionality of open source software for all levels of the quantum software chain \cite{fingerhuth2018open}.

As access to quantum computing hardware is still limited, software stacks typically include a simulator that emulates the quantum circuit execution on a classical computer. Due to the exponential scaling of memory resources with system size for these simulators, there is a limit of approximately 45-50 qubits even on the largest classical supercomputers. In some cases, the software stack can be configured to access actual quantum hardware through a cloud-computing interface where the specific hardware available depends on the provider.

While the majority of software is targeted towards general quantum programs, there are a few libraries that have been designed and built specifically for facilitating simulations of materials on quantum computers. These libraries provide interfaces that are more intuitive to scientists in the chemistry and materials domains, as well as high-level functions that allow users to work at an abstraction layer above gate-level composition of quantum circuits.  Some of the listed packages can even interface with conventional computational packages that are required to provide the one- and two-electron integrals required for simulations in quantum chemistry.  In Table~\ref{softwareTable}, we describe the noteworthy attributes of various software packages available for designing materials simulations on quantum computers.  Though some of the software packages were developed specifically for quantum chemistry, many components of them can be adapted for more general materials simulations.

\begin{table}
    \caption{\label{softwareTable}Open-source software for simulating materials on quantum computers}
    \footnotesize
    \rowcolors{2}{gray!25}{white}
    \begin{tabular}{ | m{10em} | m{2cm}| m{26em} | }
    \br
    \rowcolor{gray!50}
    \textbf{Software Package}&\textbf{Developer}&\textbf{Noteworthy Attributes}\\
	Quantum \newline{} Development Kit &  Microsoft & Quantum chemistry library for easy integration with NWChem \\ 
	OpenFermion~\cite{mcclean2020openfermion} & Google & Specialized data structures, functions, and interfaces for electronic structure calculations\\ 
	Qiskit Aqua~\cite{Qiskit} & IBM & Supports ground state energy computations, excited states and dipole moments of molecules, including open and closed-shell; integrated access to IBM quantum hardware \\
	Grove~\cite{smith2016practical} & Rigetti & Algorithms for VQE and QPE for quantum chemistry simulations; integrated access to Rigetti quantum hardware \\ 
	Pennylane~\cite{bromley2020applications} & Xanadu & Methods for VQE, thermal state preparation, and computation of molecular vibronic spectra \\ 
	XACC~\cite{mccaskey2020xacc} & ORNL & Methods for VQE, QPE, and real- and imaginary-time evolution for use in materials simulations \\
	MISTIQS~\cite{powers2021mistiqs} & USC & Full-stack, high-level programming library for dynamic simulations of the Heisenberg model\\
	ArQTiC~\cite{bassman2021arqtic} & LBNL & Full-stack, high-level programming library for zero- and finite-temperature dynamic simulations of materials modeled by spin-lattices\\
	\br
    \end{tabular}\\
\end{table}
\normalsize

\section{Algorithms for Materials Simulations}
\label{sec:algorithms}
Hundreds of quantum algorithms have been developed since the early days of quantum computing~\cite{Montanaro:2016, Jordan:2021}.  While there are myriad properties that are of interest for materials simulations on quantum computers, only a small number of quantum algorithms currently serve as building blocks for designing such quantum programs~\cite{bauer2020quantum}.  Of the relevant quantum algorithms that can be adapted to materials simulations problems, many require substantial quantum resources, like large numbers of qubits or deep quantum circuit. A large area of research is therefore dedicated to reducing the number of required qubits and gates for those algorithms relevant for chemistry and materials science.  A second large area of research involves developing novel algorithms for materials simulation.  In this section, we summarize the main algorithms that have been developed for simulations of materials on quantum computers.

\subsection{State Preparation}
\label{subsec:state_prep}
Central to materials simulations on quantum computers is the ability to prepare a target wave function $|\psi_t\rangle$ accurately on a quantum device. Consider a qubit register in its fiducial initial state $|\psi_0\rangle$, usually all spins aligned in one direction. In general, to prepare the target wave function one must apply a unitary operator $\hat{U}$ with an eigenbasis containing that wave function.  Application of $\hat{U}$ to the initial product state, results in a final wavefunction $|\psi_f\rangle$, written as
\begin{equation}
\label{Eq:Basic-state-prep}
\hat{U}|\psi_0\rangle = |\psi_f\rangle.
\end{equation}
A state preparation algorithm must be able to prepare a state where we minimize infidelity to some algorithmic $\epsilon$ of error
\begin{equation}
    1-|\langle\psi_f|\psi_t\rangle| \leq \epsilon.
\end{equation}
Once the target state has been prepared, observable information can be extracted through measurement in static materials simulations. Alternatively, the target state can represent an initial quantum state  that is further evolved through time in dynamic materials simulations.  

The preparation of an arbitrary quantum state is exponentially hard because a generic state for an $n$-qubit system contains $2^n$ complex amplitudes, thus requiring $\mathcal{O}(2^n)$ gates in a quantum circuit to prepare \cite{nielsen2010quantum}.  Fortunately, most materials problems do not begin with an arbitrary quantum state, but rather with structured states including product states, ground states, and thermal states.  In fact, if the wave function of the system is represented on a real-space grid, then most physically relevant quantum states are efficiently preparable \cite{zalka1998simulating, ward2009preparation}.  The computational complexity and resource estimates of  general-purpose ground state preparation algorithms and their computational complexity are analyzed in Refs. \cite{ge2019faster,lin2020near, lemieux2020resource}.

\subsubsection{Ground State Preparation}
A large majority of materials simulations require initializing the system into the ground state of some Hamiltonian, including static simulations aimed at computing the ground state properties of a material \cite{li2011solving} and dynamic simulations such as quantum quenches \cite{mitra2018quantum}.  There are a number of methods for preparing the ground state of a given material Hamiltonian on quantum computers.  One approach is based on the quantum phase estimation (QPE) algorithm, which can be used to find the phase $\theta$ of an eigenvalue of a unitary matrix, i.e., $\hat{U} \ket{\psi} = e^{2 \pi i \theta} \ket{\psi}$, where $\hat{U}$ is a unitary matrix and $\ket{\psi}$ is its eigenvector \cite{kitaev1996quantum, abrams1999quantum, nielsen2010quantum}.  If we set $\hat{U} = e^{-i \hat{H} t}$, where $\hat{H}$ is the system Hamiltonian, then the eigenvalues are proportional to the energy levels of the system~\cite{aspuru2005simulated}. Within materials problems, therefore, QPE is generally used to find the ground state energy of a material, which is simply the smallest eigenvalue of the material's Hamiltonian.

Two quantum registers of qubits are used in QPE. The first register is initialized into an efficient-to-prepare state that has ample overlap with the eigenstate of the corresponding desired eigenvalue.  For example, if the ground state energy is desired, this register should be initialized into a state $|\psi\rangle$ that has sufficient overlap with the ground state $|g\rangle$.  The second register is initialized into an equal superposition over all computational basis states and at the end of the algorithm will store the desired eigenvalue in binary format with probability $|\langle \psi |g\rangle|$.  Thus, the higher the overlap between $|\psi\rangle$  and $|g\rangle$, the higher the probability of successfully measuring the desired eigenvalue.  After measurement of the eigenvalue into the second register, the first register collapses into the corresponding eigenstate.  Thus, QPE can also be utilized for ground state preparation, where the first register, storing the eigenstate, is used as the initial state for a materials simulation problem.

The difficulty with this method is that initializing the quantum register into a state with sufficiently large overlap with the desired eigenstate (usually the ground state) can be a nontrivial task. Typically, therefore, one must perform the QPE repeatedly until the ground state is obtained.  This incurs a computational cost that scales inversely with the size of the overlap between the initial and ground state, though techniques have been proposed to reduce this cost \cite{svore2013faster, berry2018improved, kivlichan2019phase}.  Techniques proposed for preparing a state close to the ground state include methods that utilize classically tractable computations \cite{wang2009efficient, babbush2015chemical, sugisaki2016quantum, sugisaki2018quantum, sugisaki2019open, tubman2018postponing}, variational methods \cite{yung2014transistor}, utilizing imaginary time evolution \cite{motta2020determining}, and adiabatic state preparation (ASP) \cite{aspuru2005simulated}.

In fact, ASP can be used in its own right to prepare ground states for static and dynamic simulations.  ASP is based on the adiabatic theorem \cite{born1928beweis}, which states that the system remains in its instantaneous eigenstate despite a slowly changing system Hamiltonian as long as there is a gap between the corresponding eigenvalue and the rest of the Hamiltonian spectrum. ASP for ground state preparation works by initializing the qubits into an easy-to-prepare ground state of some initial Hamiltonian and then slowly varying the Hamiltonian into a final Hamiltonian whose ground state is the target ground state. The speed of this variation is limited by the size of the gap between the energies of the ground state and the first excited state. 

A final method for ground state preparation is the variational quantum eigensolver (VQE)~\cite{peruzzo2014variational, mcclean2016theory}.  VQE is a hybrid quantum-classical algorithm in which a parameterized quantum circuit is used to construct a wavefunction while a classical computer is used to optimize these parameters to minimize the expectation value of the Hamiltonian.  VQE can be summarized in the following steps: (i) start with a random set of parameters $\theta$, (ii) prepare the trail wavefunction $\ket{\psi\left(\theta\right)}$ on the quantum computer, (iii) measure the expectation value of the Hamiltonian for $\ket{\psi\left(\theta\right)}$, (iv) find a new set of parameters $\theta$, (v) repeat until the convergence in energy is achieved. At this point, the parameterized circuit should prepare the ground state, or a state very close to the ground state, of the Hamiltonian.  VQE requires substantially smaller number of gates and shorter coherence times than QPE. It trades a reduction in required coherence time with a polynomial number of repetitions. It is thus better suited for NISQ architectures.

\subsubsection{Thermal States Preparation}
While many state preparation techniques assume the system to be at zero temperature, initializing the system into a thermal state is required for interesting materials simulations related to thermalization, thermal rate constants, and other finite-temperature phenomena.  Various methods for thermal state preparation have been proposed, including one for preparing Gibbs states that makes use of QPE \cite{riera2012thermalization}, methods that rely on quantum imaginary time-evolution \cite{motta2020determining, sun2020quantum}, and methods that prepare thermofield double states \cite{martyn2019product, cottrell2019build, wu2019variational, zhu2020generation}, many of which are inspired by the quantum approximate optimization algorithm (QAOA) \cite{farhi2014quantum}.

\subsection{Hamiltonian evolution}
\label{subsec:ham_evo}
After preparing the initial state of the system, most materials simulations require evolving the system through time, defined by the Schrodinger equation
\begin{equation} \label{schrodinger}
    i \hbar \frac{\partial}{\partial t} \ket{\psi(t)} = \hat{H} \ket{\psi(t)}
\end{equation}
where $\hat{H}$ is the simulated system's Hamiltonian, and $\ket{\psi(t)}$ is the time-dependent many-body wavefunction of the system.  The time-evolution operator, which acts on a system's wavefunction, taking it from its form at the initial time to its form at a final time, is derived from the time-dependent Schr\"{o}dinger equation, and can be written in the atomic unit as a time-ordered exponential 
\begin{equation} \label{evolution_operator}
\hat{U}(0,t) \equiv \hat{U}(t)= \mathcal{T} \exp\left[-i\int_{0}^{t}\hat{H}(t)dt\right]
\end{equation}
For a time-independent Hamiltonian, equation (\ref{evolution_operator}) simplifies to $\hat{U}(t) = e^{-i\hat{H}t}$.  For a time-dependent Hamiltonian, a product formula is used to cut the time-evolution into small time-steps over which period the Hamiltonian is assumed to be constant \cite{poulin2011quantum}.

In order the simulate such Hamiltonian dynamics on a quantum computer, a quantum circuit must be generated whose comprising gates mimic the operation of the time-evolution operator on the qubits.  Unfortunately, the exact time-evolution operator can be difficult to compute in many cases, as it is given by an exponentiated matrix (the system Hamiltonian).  Specifically, if the Hamiltonian operator is difficult to diagonalize, the time-evolution operator will be difficult to compute.  For some special models, there are procedures for diagonalizing the system Hamiltonian.  In these cases, the quantum circuit can be built up from (i) a set of gates which move the qubits into the diagonal basis, (ii) a set of gates that enact the time-evolution operator in its easily computed diagonal form, and (iii) a set of gates that revert back to the computational basis of the qubits \cite{verstraete2009quantum, cervera2018exact}.  However, such procedures are not known for the majority of material systems, and so approximations are generally made for computing the time-evolution operator.

The simplest and most commonly used approximation is based on the Trotter--Suzuki decomposition \cite{trotter1959on, suzuki1976generalized}.  First, the Hamiltonian is split up into terms that are each easily diagonalizable on their own.  For example, consider a Hamiltonian $\hat{H}$ that can be decomposed into two separate components as $\hat{H} = \hat{X} + \hat{Y}$, where $\hat{X}$ and $\hat{Y}$ are easy to diagonlize individually.  The time-evolution operator $\hat{U}= e^{-i\hat{H}t} = e^{-i(\hat{X} + \hat{Y})t}$ requires exponentiation of this Hamiltonian; however, if $\hat{X}$ and $\hat{Y}$ do not commute, the exponential law $e^{X+Y} = e^{X}e^{Y}$ is invalid.  Generally, Hamiltonians for materials systems are composed of linear combinations of operators that do not commute.  However, applying the Lie--Trotter product formula
\begin{equation}
\label{Lie-Trotter}
    e^{-i(\hat{X}+\hat{Y})t}= \lim_{N\rightarrow \infty} \left(e^{-i\hat{X}t/N}e^{-i\hat{Y}t/N}\right)^N
\end{equation}
it is possible to approximate the exponential in the time-evolution operator.  In this way, the time evolution is broken into small time-steps $\Delta t = \frac{t}{N}$.  For each time-step the individual terms of the Hamiltonian are exponentiated and the matrix exponentials are then multiplied together.  The evolution by different components of the Hamiltonian are each alternatingly applied to the system for a short time until the total time evolution is reached.   This method (commonly known as Trotterizing) makes time-evolution of materials on quantum computers feasible.  Since in practice $N$ in equation (\ref{Lie-Trotter}) is finite, the Trotter approximation leads to an amount of error controlled by the size of the time-step $\Delta t$.  For a detailed discussion on Trotter error in quantum algorithms see Refs.~\cite{whitfield2011simulation, childs2019trotter}.  Less commonly used approximations are based on a Taylor series truncation for time-independent Hamiltonians \cite{berry2015simulating}, and the Dyson series for time-dependent Hamiltonians \cite{kieferova2019simulating}.  In both cases, the time-evolution operator is built up as a linear combination of unitaries \cite{childs2012hamiltonian}.  These methods, however, require high quantum resources, and thus are not as frequently used in the NISQ era.

When evolving the system through real-time, the time-evolution operator is unitary (by definition), and thus it is relatively straightforward to convert it into a set of gates in a quantum circuit that can be executed on the quantum computer.  However, it is also possible to construct evolution through "imaginary time".  This is accomplished by setting the normally real-valued $t$ in $\hat{U}(t) = e^{-i\hat{H}t}$ to a purely imaginary value $t = -i\beta$, such that the evolution operator becomes $\hat{U}(t) = e^{-\beta\hat{H}}$.  This non-unitary operator can be useful for several purposes, including driving a system into Hamiltonian ground- and excited-states, or creating thermal states \cite{motta2020determining, mcardle2019variational, white2009minimally}.  Since it is not possible to implement non-unitary gates on the qubits of a quantum computer, this operator must be approximated using the recently developed quantum imaginary time-evolution (QITE) method \cite{motta2020determining}.  While theoretically sound, QITE can produce relatively large circuits which are not NISQ-friendly, and thus several techniques for shortening circuit depths have been presented \cite{sun2020quantum,  yeter2020practical, gomes2020efficient}.


\subsection{Embedding Methods}
\label{subsec:embedded}
\begin{figure}
    \centering
    \includegraphics[scale=0.55]{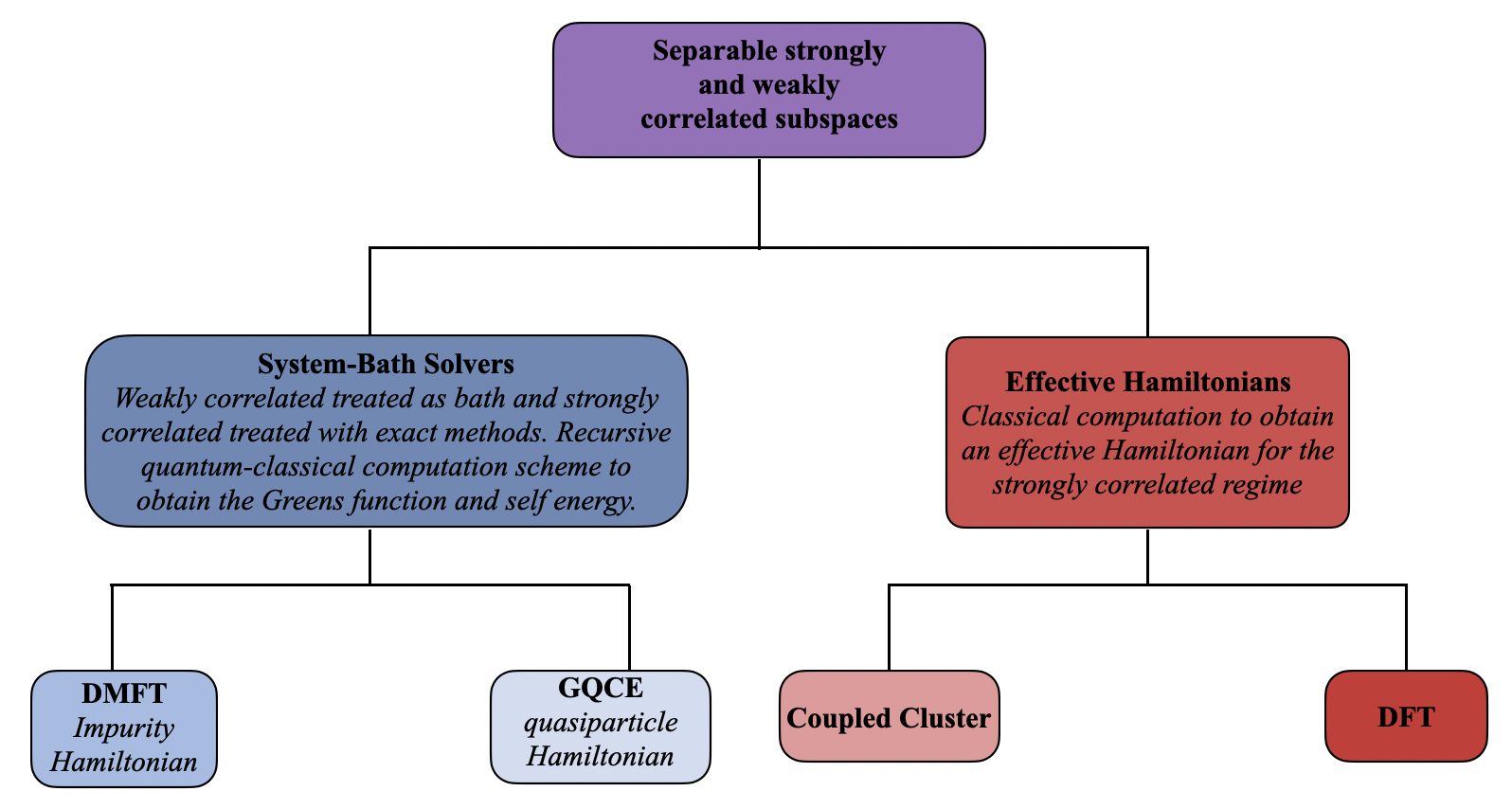}
    \caption{Embedding schemes for hybrid quantum-classical computation of materials.}
    \label{fig:embedding}
\end{figure}

Quantum processing power for materials simulations is maximized for strongly correlated systems where perturbative approaches are incapable of capturing correlation energy and dynamical effects. In many relevant materials only a small portion of the system is strongly correlated (usually this is the main region of interest), while the larger remaining portion can be considered weakly correlated.  High-accuracy calculations are required to characterize this strongly correlated region, while lower-accuracy calculations are often sufficient to characterize the weakly correlated region.  Quantum embedding theories seek to join these two levels of computation together to accurately simulate larger quantum materials with minimal computational resources.  This is achieved by embedding the strongly correlated space into the remaining weakly correlated space, where the strongly correlated component is treated with a high-accuracy calculation techniques and the weakly correlated component is calculated using cheaper or more approximate computational methods~\cite{SunRev2016, jones2020embedding}.  Near-term quantum computers, which can efficiently simulate strongly correlated systems but have a limited number of qubits, are well-suited for the high-accuracy calculation of the reduced system space.  Classical computers, on the other hand, are well-suited for simulation of the larger, weakly correlated space, for which they can use a host of approximation schemes that have been developed over the last few decades. Therefore, the development of hybrid quantum-classical embedding algorithms offers great potential in enabling simulations of strongly correlated materials in the near term.

Hybrid quantum-classical embedding algorithms can be divided into two main groups: (i) system-bath solvers that encode weakly correlated degrees of freedom as a bath (i.e., environment) that is connected to the strongly-correlated subsystem, and (ii) downfolding contributions from high, unoccupied states into an effective Hamiltonian in a reduced Hilbert space.  A schematic categorizing the various embedding methods is depicted in Figure \ref{fig:embedding}. 

System-bath solvers use a hybridization function to converge an impurity Green's function to a lattice Green's function. Traditionally, the impurity Hamiltonian, which represents the strongly correlated subsystem, is solved using an exact solver.  The complexity of this problem scales exponentially with subsystem size on classical computers, which makes using a quantum computer to solve this portion of the embedding problem attractive.  Thus hybrid classical-quantum embedding algorithms can be devised that use a feedback loop between a quantum computer, which computes the impurity Green's function and a classical computer, which computes the lattice Green's function, to achieve convergence between the two. System-bath embedding algorithms used to evaluate strongly correlated material systems are Dynamical Mean Field Theory (DMFT)~\cite{Zgid2011}, Density Matrix Embedding Theory (DMET)~\cite{Chan_DMET}, and Self-Energy Embedding Theory (SEET)~\cite{Zgid_SEET}.  Since these methods rely on results from a quantum computer to achieve convergence, advances in quantum algorithms that can calculate the Green's function and self-energy in a noisy system are essential. 

The principal DMFT algorithm for quantum computers uses real-time evolution and the QPE to obtain the impurity Green's function for the subsystem \cite{maier2005quantum, bauer2016hybrid, rubin2016hybrid, kreula2016few, keen2020quantum, Jaderberg2020}.  Due to noise on current quantum hardware, the unitary time-evolution required to obtain necessary correlation functions becomes difficult to accurately simulate since the decoherence time of the qubits is shorter than the amount of time needed to prepare the time-evolved state.  As an alternative, variational methods can be used for near-term quantum devices, where the ground state is prepared on the quantum device using a variational algorithm and the impurity Green's function is extracted using the Lehman representation in the zero temperature limit~\cite{rungger2019dynamical}.  While the variational and QPE DMFT algorithms are formally equivalent in solution, they have very different noise sensitivities and scaling. The variational methods currently rely on VQE, where the complexity scaling to larger system sizes is not known, and errors arise in the energies obtained.  The distribution of energies should obey sum rules and causal relations, neither of which are guaranteed by the quantum algorithm, and thus some form of regularization is required to impose these.  Finally, the real-time evolution leads to Trotter errors, which affect the measured energies.

Errors from the measurements in both variational and QPE DMFT approaches will propagate through the self-consistency loop (although so far this has not yet been demonstrated in the variational approach), leading to unphysical quasiparticle weights and unphysical poles near zero energy. Empirically, it is observed that these issues are more severe in the metallic phase than in the insulating phase \cite{keen2020quantum}, making a converged solution for metallic materials difficult to obtain. The dependence of the self-consistent algorithm on the result of the impurity solver, which here is obtained from a quantum computer, makes converging these embedding methods noise sensitive.

An alternative approach to DMFT, called the Gutzwiller Quantum Classical Embedding (GQCE) algorithm, defines a strongly-correlated subsystem interacting with a quasi-particle bath~\cite{Yao2020}. Rather than evaluating the Green's function and self-energy to reach self-consistency, GQCE only needs the ground state and the single particle density matrix of the embedded subsystem which greatly reduces the complication of extracting the full Green's function from a quantum computer. 

Unlike the system-bath solvers, embedding in method (ii) occurs as a classical pre-processing technique.  Specifically, the Hilbert space of a large system is embedded into a reduced Hilbert space, approximately maintaining all correlation contributions from the initial Hilbert space.  Typically, the basis set used to define a material system includes higher-energy, unoccupied states, which contain dynamical correlation. Encoding a Hamiltonian using such an extended basis set requires a larger number of qubits to capture the effects of these orbitals. If one were to evaluate this Hamiltonian classically, approximate solvers are required to extract the energy. To use an exact solver on a classical computer, the Hilbert space must be truncated and thus, unoccupied orbital contributions are lost. Embedding methods using density functional theory (DFT) and the double unitary coupled cluster (DUCC) have been developed to capture correlations contained in the virtual (i.e., unoccupied) orbitals by downfolding contributions from these orbitals into a reduced-space, effective Hamiltonian ~\cite{bauman2019downfolding, Metcalf2020, ma2020quantum}.  To obtain the effective Hamiltonian the virtual orbitals containing dynamic correlation are separated from the subspace containing static correlation effects. Projecting the virtual space (environment) onto the reduced active space yields an effective Hamiltonian that provides higher accuracy without the need to model the entire orbital space, thus requiring fewer computational resources. DFT and DUCC calculations are performed on a classical computer and the resulting effective Hamiltonian in a reduced space is evaluated on the quantum computer using algorithms like VQE or QPE.

\section{Qubit Representation}
\label{sec:qubit_rep}
The first step of quantum simulation is choosing a representation for the system such that it can be stored on and manipulated by a quantum computer.  There are a variety of methods for modeling physical systems and mapping these models onto the qubits of a quantum computer \cite{kassal2011simulating, mcardle2020quantum}.  In materials simulations, the system of interest could be a free-standing composition of atoms such as simple molecules, proteins, or polymers, which are generally seen in the context of quantum chemistry \cite{cramer2013essentials}; alternatively the system could be a crystalline structure represented by atoms at fixed points in a unit cell, which is repeated in all directions, as is generally seen in the context of materials science \cite{ashcroft1976solid}; and finally, the system can be represented by a spin-lattice model, characterized by spins residing on, and possibly hopping between, points on a lattice, an approach generally taken in condensed matter physics \cite{parkinson2010introduction}.  Based on the representation of the quantum material, the Hamiltonian of the system and how system states are mapped to the qubits of a quantum computer can be carried out in various ways.  

\begin{figure}
    \centering
    \includegraphics[scale=0.55]{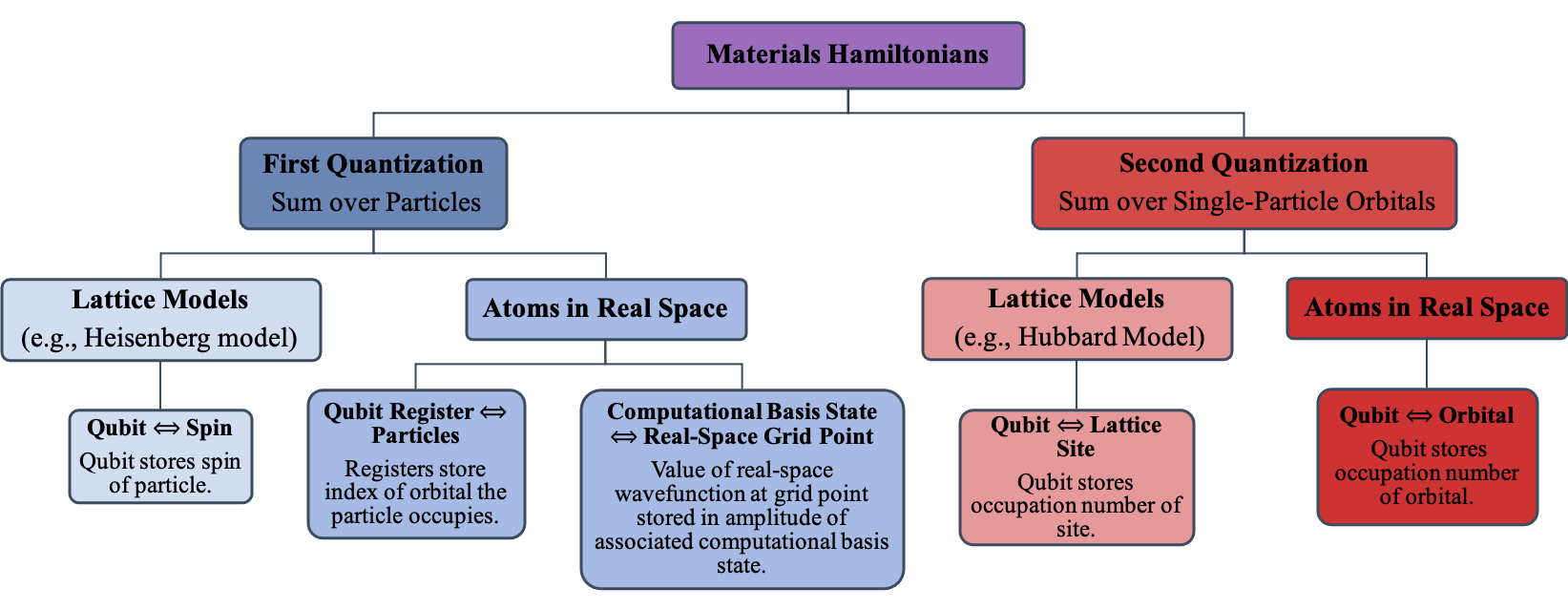}
    \caption{Tree diagram for Hamiltonians and qubit mappings that have been used for simulating quantum materials on quantum computers.}
    \label{hamTree}
\end{figure}

Figure \ref{hamTree} presents a tree diagram for various Hamiltonians and qubit mappings that have been used to model material systems for simulation on quantum computers.  At the top level, one decides whether to write the Hamiltonian in the first or second quantization \cite{kassal2011simulating}.  The main difference between the first and second quantizations is whether the antisymmetry of the wavefunction (necessary for simulating fermions, which comprise most materials) is dealt with in the composition of the initial wave function (first quantization), or in the construction of the operators acting on the wave function (second quantization).  These Hamiltonian formulations carry different complexity scaling with respect to the size of the system and its state space, so the specifics of the simulation problem and available quantum resources dictate which modeling paradigm should be chosen.  After choosing a quantization, one must choose whether to represent the material exactly in terms of the positions and species of the comprising atoms or to abstract away important parameters of the system into a lattice-type spin model.  Again, specifics of the simulated material and the desired observable dictate which approach is best to use.  In the subsections below, we review each of the models in greater detail.

\subsection{First Quantization}
In the first quantization, the Hamiltonian is defined by sums over the particles in the system.  In this sense, particles are distinguishable, and care must be taken to anti-symmeterize initial states when simulating fermions \cite{abrams1997simulation}.  Within the first quantization, materials can be modeled by spin-lattice models or defined by the real-space positions of their constituent atoms.  The two most commonly used spin-lattice models in the first quantization are the Heisenberg model and the Ising model, which is actually a subset of the Heisenberg model.

The general Heisenberg model Hamiltonian is given by 
\begin{equation}
H(t) =  -J_{x}\sum_{\langle i,j \rangle} \sigma_{i}^{x}\sigma_{j}^{x} -J_{y}\sum_{\langle i,j \rangle} \sigma_{i}^{y}\sigma_{j}^{y} -J_{z}\sum_{\langle i,j \rangle} \sigma_{i}^{z}\sigma_{j}^{z} - h_{z}(t) \sum_{i=1}^{N} \sigma_{i}^{z}
\label{Hamiltonian}
\end{equation}
where $\sigma_{\alpha}$ are the Pauli matrices for $\alpha = {x,y,z}$; $J_x, J_y, J_z$ are the strengths of the exchange interactions between pairs of particles $\langle i,j \rangle$ in the $x-$, $y-$, and $z-$directions, respectively; $h_z$ is the strength of an external magnetic field (which without loss of generality is assumed to be in the $z$-direction); and $N$ is the number of spins.  Popular derivatives of this general model include the transverse field Ising model where $J_z = J_y = 0$ \cite{pfeuty1970one, stinchcombe1973ising}, the XY model where $J_z = 0$ \cite{lieb1961two, barouch1970statistical, toner1995long}, and the XXZ model where $J_x = J_y = J$ \cite{orbach1958linear, des1966anisotropic, kubo1988existence}.  These models represent spins on a fixed lattice with exchange interactions between pairs of spins (usually nearest neighbors), in the presence of an external magnetic field.  These models can be used to study critical points, phase transitions, transport, and entanglement in magnetic materials \cite{kosterlitz1974critical, calabrese2011quantum, heyl2013dynamical, narozhny1998transport, wang2001entanglement, gu2005ground, gu2006local}.  For these lattice models, there is a one-to-one mapping between spins in the model and qubits on the quantum computer.  

For simulating atoms in real-space, the first quantized form of the Hamiltonian for a general system is given by
\begin{equation} \label{first_quant_ham}
    H = \sum_{i=1}^{N} \frac{p_{i}^{2}}{2m_i} + U(x_1, ..., x_N)
\end{equation}
where $N$ is the number of particles in the system, $p_i$ and $m_i$ are the momentum and mass of particle $i$, respectively, and $U$ is the potential energy of the many-body system.  For Hamiltonian (\ref{first_quant_ham}), there are two different ways to map the problem onto qubits: (i) a single-particle basis set method \cite{abrams1997simulation} and (ii) a real-space grid-based method \cite{wiesner1996simulations, zalka1998simulating}.  In the single-particle basis set method, a discrete single-particle basis (e.g., molecular orbitals or planewaves) is constructed for the many-body wavefunction representing the material. The $M$ basis states are assigned an integer number from $0$ to $M-1$ for indexing, which can be stored in $\lceil \log_2 M \rceil$ qubits.  Thus, $\lceil \log_2 M \rceil$ qubits are then grouped into a quantum register, and one quantum register is prepared for each of the $N$ particles enumerated in Hamiltonian (\ref{first_quant_ham}).  Each quantum register stores the index of the single-particle basis state that its corresponding particle occupies.  In this method, care must be taken to anti-symmetrize the initial wavefuction \cite{berry2018improved}. A variety of work has studied the computational complexity of algorithms using different basis sets for this method of mapping of the material system onto qubits \cite{toloui2013quantum, babbush2017exponentially, babbush2018low, babbush2019quantum}.

In the real-space grid-based method, the many-body wavefunction of the quantum material is defined on a discretized real-space grid.  For a single particle in one dimension, space can be discretized into $2^m$ points.  Each of these points can be mapped to one of the computational basis states of an $m$-qubit system.  The wavefunction is expanded in the computational basis of the multi-qubit system as $|\psi\rangle = \sum_{x=0}^{2^m-1}a_x|x\rangle$, where each of the computational basis states $|x\rangle$ correspond to one of the spatial grid points.  The value of the real-space wavefunction at a given grid-point is stored in the amplitude of the corresponding basis state.  This scheme can easily be extended to $N$ particles in $d$ dimensions using $Ndm$ qubits. A few works have analyzed algorithms using grid-based methods \cite{kassal2008polynomial, omalley2016scalable, kivlichan2017bounding}.

\subsection{Second Quantization}
In the second quantization, the Hamiltonian is defined by sums over basis states.  In this regime, particles are indistinguishable, as it is only necessary to keep track of the number of particles that occupy a given state (0 or 1 for fermions).  Thus, care must be taken to ensure that the operators acting on the distinguishable qubits are anti-symmetric.  Various transformations, including the Jordan-Wigner \cite{jordan1993paulische}, Bravyi-Kitaev \cite{bravyi2002fermionic}, and others \cite{ball2005fermions, verstraete2005mapping, seeley2012bravyi, farrelly2014causal, havlivcek2017operator, jiang2019majorana, steudtner2019quantum, setia2019superfast} have been developed for this.  As in the first quantization, in the second quantization materials can be modeled by spin-lattice models or defined by the real-space positions of their constituent atoms.  

The most commonly used spin-lattice model in the second quantization is the Hubbard model, a minimum model that accounts for the quantum mechanical motion of spins on a lattice as well as the repulsive interaction between them \cite{tasaki2020introduction}.  It is defined as 
\begin{equation} \label{hubbard}
    H = -\sum_{\langle i,j \rangle} \sum_\sigma t_{ij} (c_{i,\sigma}^{\dagger}c_{j,\sigma} + c_{j,\sigma}^{\dagger}c_{i,\sigma}) + \sum_{i} U_i n_{i,\uparrow}n_{i,\downarrow}
\end{equation}
where $t_{ij}$ is the hopping term between pairs of nearest neighbor sites $\langle i,j \rangle$, $c_{i,\sigma}^{\dagger}$($c_{i,\sigma}$) is the creation (annihilation) operator for a particle on site $i$ of spin $\sigma$, $U_i$ is the on-site interaction term for lattice site $i$, and $n_{i,\sigma} = c_{i,\sigma}^{\dagger}c_{i,\sigma}$ is the number operator, which counts how many particles of spin $\sigma$ occupy site $i$.  Commonly used in condensed matter physics, the Hubbard model has been shown to exhibit myriad physical phenomena including metal-insulator transitions \cite{bulla1999metal}, ferromagnetism \cite{mielke1993ferromagnetism}, antiferromagnetism \cite{hirsch1989antiferromagnetism}, and superconductivity \cite{maier2000d}.  It can also serve as a radically simplified version of the electronic structure Hamiltonian \cite{wecker2015solving}.

For lattice models in the second quantization, there is a one-to-one mapping between lattice spin-orbitals in the model and qubits on the quantum computer.  The value of the qubit in the computational basis corresponds to the occupation number of the lattice spin-orbital; for example, a measured qubit value of of '0' represents an unoccupied spin-orbital, while a '1' represents an occupied spin-orbital.

The second quantized form of the Hamiltonian for simulating atoms in real-space is given by
\begin{equation} \label{second_hamiltonian}
    H = \sum_{pq} h_{pq} c_p^{\dagger}c_q + \sum_{pqrs} h_{pqrs} c_p^{\dagger} c_q^{\dagger} c_r c_s
\end{equation}
where $p,q,r,s$ index the basis functions used (e.g., planewaves, Gaussian orbitals, etc.) and $c_i^{\dagger}$ ($c_i$) are the creation (annihilation) operators that create (destroy) a particle in the basis state $i$. The coefficients $h_{pq}$ are one-particle integrals involving the kinetic energy and background potential energy, while $h_{pqrs}$ are two-particle integrals involving interaction energies between the particles \cite{whitfield2011simulation}.  There is a one-to-one mapping between spin-orbitals and qubits, where the value of the qubit in the computational basis represents the occupation number of the associated spin-orbital.  A number of works have proposed algorithms using the second quantization representation and associated computational complexities \cite{mcclean2014exploiting, moll2016optimizing, babbush2016exponentially, babbush2018low, motta2018low, low2018hamiltonian, mcclean2020discontinuous, takeshita2020increasing}.

\section{Static Material Simulations}
\label{sec:static}
Quantum computers are physical devices that behave according to the laws of quantum mechanics and thus, can simulate the dynamics of other quantum systems.  This makes simulating the time evolution of a quantum system a natural task for a quantum computer. Algorithms for analyzing static properties of a physical system, i.e., finding its ground state or low-lying excited states, are more complex. This section summarizes efforts to calculate static material properties on NISQ computers.

While small quantum computers already exist and can be used to solve scientific problems, they are still too limited to provide a practical advantage over classical computers. The studied problems are mostly toy problems that can be easily solved on classical computers. To our knowledge, the largest calculation of static properties used 18 qubits~\cite{aleiner2020accurately}. In the domain of materials science, current efforts are focused on studying properties of simple quantum models that capture essential material properties, e.g., the Ising, Heisenberg, Hubbard, or similar models. These models capture the electronic structure in solids. The goal of current efforts is therefore to find the electronic ground or low-lying states, thermal states, or calculate other electronic properties.  The prospects of scaling experiments to tens of qubits makes solving problems beyond the reach of classical computers an exciting possibility in near future.

This review covers only digital quantum computers. There are other approaches for finding the static properties of quantum systems. In particular, analog quantum simulators~\cite{georgescu2014quantum, gross2017quantum}, typically implemented in optical lattices, can be used to directly create a physical system that approximates a given Hamiltonian. Similarly, quantum annealers~\cite{das2008colloquium, albash2018adiabatic, hauke2020perspectives} utilize the adiabatic theorem to prepare a ground state of a physical system implementing a given quantum model. The disadvantage of these other implementations is that the set of models accessible to these methods is restricted by their physical limitations. The advantage of digital quantum computers is therefore their universal programmability and applicability to wide spectrum of scientific problems.

\subsection{Ground States}

Calculating ground state properties of quantum systems is a fundamental problem in quantum mechanics. The energies of the ground state and first few excited states, expectation values of various operators, and correlation functions are the typical quantities of interest. Usually, the Hamiltonian depends on a set of parameters, and the ground state properties are calculated as functions of these parameters.

The QPE algorithm~\cite{kitaev1996quantum, nielsen2010quantum} is a well-known algorithm for calculating the ground state energy on a quantum computer, as discussed in Section \ref{subsec:state_prep}. It requires preparation of a trial state that has an overlap with the ground state. The trail state is then evolved using a controlled evolution and the lowest eigenvalue of the Hamiltonian is estimated by performing the quantum Fourier transform. The algorithm requires deep circuits to approximate the controlled evolution with sufficient accuracy. Due to this requirement, QPE is not very practical on current NISQ devices and is better suited for future fault-tolerant quantum computers. QPE has been experimentally demonstrated for the quantum spin models in Refs.~\cite{li2011solving, cruz2020optimizing}.

The VQE algorithm, as discussed in Section \ref{subsec:state_prep}, has been developed to overcome these issues. It is a hybrid quantum-classical approach where a quantum computer is used to prepare and measure a parametrized ansatz and a classical computer is used to find parameter values that minimize the calculated energy. The method has been demonstrated experimentally on various problems in quantum chemistry~\cite{peruzzo2014variational, omalley2016scalable, kandala2017hardware, shen2017quantum, hempel2018quantum, colless2018computation, nam2020ground}. It has been shown that VQE is fairly robust to errors present in NISQ devices.

In the context of materials science, VQE has been used to study the Heisenberg antiferromagnetic model~\cite{kandala2017hardware}, the four-site Hubbard model with half-filling~\cite{xu2020test}, and the Hubbard model restricted to a subspace with a fixed number of electrons~\cite{montanaro2020compressed}. Several ansatz variants have been analyzed in Refs.~\cite{dallaire2019low, sokolov2020quantum, yoshioka2020variational}, and the performance and required resources for strongly-correlated systems in Refs.~\cite{jiang2018quantum, cai2020resource, uvarov2020variational, manrique2021momentumspace}. An alternative approach is to use the variational Hamiltonian ansatz inspired by the adiabatic time-evolution operator~\cite{wecker2015progress, wiersema2020exploring}. Its viability for quantum computers with imperfect gates was analyzed on the Hubbard model~\cite{reiner2019finding} and the method has been further extended to find ground states with broken symmetries~\cite{vogt2020preparing}. Finally, a comprehensive analysis of variational algorithms for finding the ground state of the Hubbard model on quantum computers was presented in Ref.~\cite{cade2020strategies}. These results demonstrate that VQE works fairly well for simple condensed matter models on NISQ computers.

Other algorithms besides QPE and VQE have emerged as well. The authors of Ref.~\cite{wecker2015solving} presented all steps necessary to find the ground state of the Hubbard model using an adiabatic evolution from prepared mean-field states. Adiabatic evolution for the Ising model was demonstrated in Ref.~\cite{hebenstreit2017compressed}. A hybrid quantum-classical approach based on classical embedding algorithms and DMFT have been developed in Refs.~\cite{rungger2019dynamical, keen2020quantum}. Authors of Ref.~\cite{ma2020quantum} developed a quantum embedding theory for calculation of strongly-correlated electronic states of active regions with the rest of the system described by the density functional theory. Another approach used the QITE algorithm, as discussed in Section \ref{subsec:ham_evo}, to find the ground state of the transverse field Ising model~\cite{motta2020determining}. An inverse power iteration technique for quantum computers has been numerically demonstrated on the Bose--Hubbard model~\cite{kyriienko2020quantum}.

While quantum computers can capture strongly-correlated states more efficiently than classical computers, the problem of finding the ground state of a quantum model is a generally a hard problem even for quantum computers. For example, the Hubbard model with local magnetic fields is QMA-complete \cite{schuch2009computational}, so the existence of an efficient quantum algorithm for this model is highly unlikely.

\subsection{Excited States}

The energy of excited states is even more important than the energy of the ground state. Authors of Ref.~\cite{ollitrault2020quantum} used the quantum equation of motion for computing molecular excitation energies. The method has been experimentally demonstrated by computing excited states of phenylsulfonyl-carbazole compounds~\cite{gao2020applications}.

\subsection{Thermal States}

The classical Metropolis sampling algorithm has been generalized to quantum the domain in Refs.~\cite{terhal2000problem, temme2011quantum, yung2012quantum}. The method can prepare thermal states of both classical and quantum models and offers a quadratic speedup over classical algorithms. It can be used to study quantum systems at arbitrary temperature. Another quantum algorithm to study thermodynamic properties of the Hubbard model was developed in Ref.~\cite{dallaire2016method}. The method uses a variation of the QPE on a quantum computer to find the Green's function of the system. An approach that uses a small set of pure states to obtain properties of a thermal state was numerically applied to a ten-site Hubbard model~\cite{cohn2020minimal}.

\subsection{Other Properties}

A major advantage of digital quantum computers is their ability to calculate various quantities that are not easily accessible in ordinary experiments. An algorithm to calculate the R\'{e}nyi entropy of a many-body system~\cite{johri2017entanglement} was experimentally demonstrated on a two-site Hubbard model~\cite{linke2018measuring}. Variational algorithm to find the entanglement spectrum of the Heisenberg model was demonstrated in Ref.~\cite{larose2019variational}. The authors of Ref.~\cite{murta2020berry} designed an algorithm for estimating the Berry phase that can be used to classify the topological order of quantum systems. Crossing a topological phase transition has been experimentally demonstrated with superconducting qubits~\cite{smith2020crossing}. Another topological application, the simulation of Majorana fermions, has been demonstrated in Ref.~\cite{xiao2020topological}. The authors of Ref.~\cite{macridin2018electron} extended fermionic quantum simulations with a phonon model and implemented their algorithm based on quantum phase estimation for the two-site Holstein polaron problem on a simulator. Two methods to calculate the Green's function of a many-body system were introduced in Ref.~\cite{endo2020calculation}. Band structures have been calculated on quantum computers in Refs.~\cite{aleiner2020accurately, cerasoli2020quantum}. Finally, the partition function zeros of a finite temperature Heisenberg model were computed~\cite{francis2020many} which can be used to calculate the free energy.

\section{Dynamic Material Simulations}
\label{sec:dynamic}
\begin{figure*}
    \centering
    \includegraphics[scale=0.7]{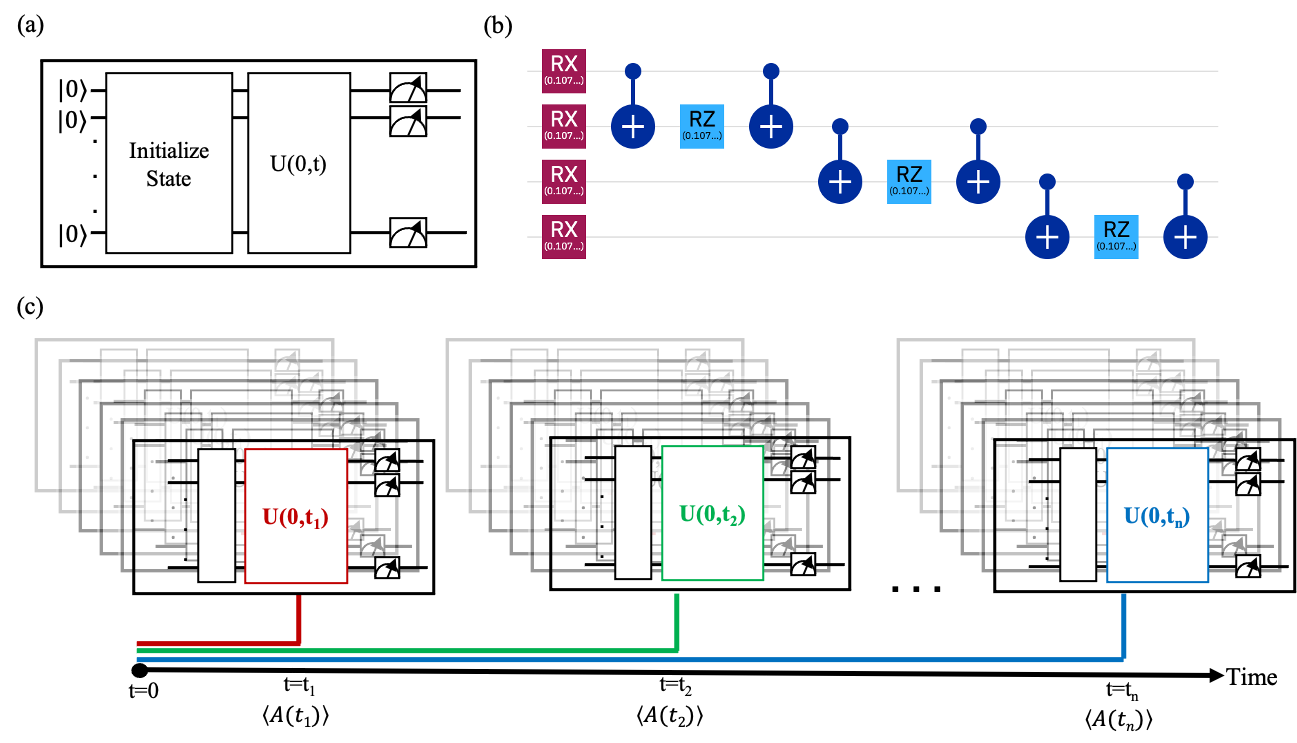}
    \caption{Quantum circuit diagrams for dynamic simulations of materials.  (a) A high-level circuit for simulating evolution for a time $t$.  After initializing the state, the time-evolution operator is applied to all qubits for a time $t$, before finally measuring all qubits. (b) A sample gate-level circuit for simulating one time-step of a 4-qubit transverse field Ising model. (c) Schematic depicting the workflow for a dynamic simulation of a material.  A different circuit must be composed for each time-step $i$, which each simulate evolution of the system from time $t=0$ to time $t_i$.  The circuit for each time-step must be run a large number of times to compute the expectation value of the desired observable $A$.}
    \label{dynamic_sims}
\end{figure*}
Measurements of quantum systems destroy the coherence of the quantum state, collapsing the many-body wavefunction of the system into an eigenstate of the measurement operator.  For this reason, dynamic simulations require a separate simulation (i.e., a separate quantum circuit) for each time-step, unless some clever weak-measurement approach is taken.  A schematic description of the general approach to dynamic simulations of materials on quantum computers is shown in Figure \ref{dynamic_sims}.  Figure \ref{dynamic_sims}a(b) shows a high-level (low-level) quantum circuit diagram for simulating time-evolution of a quantum system.

In quantum circuit diagrams, the horizontal lines represent different qubits in the system, and blocks on top of these lines represent quantum gates acting upon those qubits.  Moving from left to right in the diagram corresponds to moving forward in processor time, thus gates are chronologically ordered from left to right.  The gauge icons at the end of the qubit lines represent qubit measurement.   Figure \ref{dynamic_sims}a shows the basic quantum circuit for simulating the evolution of a system from time $t=0$ to a time $t$.  It begins by initializing the system to the desired initial state, then applies the time-evolution operator for a time $t$, and finally measures all the qubits in the system.  Figure \ref{dynamic_sims}b gives an example of a more detailed circuit diagram, splitting the high-level boxes of Figure \ref{dynamic_sims}a into one- and two-qubit gates that can be performed on quantum hardware.  Specifically, it shows one time-step of evolution for a four-qubit transverse field Ising model.

Figure \ref{dynamic_sims}c shows the workflow for carrying out dynamic simulations on quantum computers.  A different quantum circuit is built for each time-step, which simulates the evolution of the system from time $t=0$ to a time $t_i$, where $i$ runs over all time-steps of the simulation.  A large number of each of these circuits must be executed to collect statistics to calculate the expectation value of the desired observable $A$.  This workflow makes it relatively complex and expensive to perform long-time dynamic simulations since many circuits must each be executed many times.  Furthermore, since circuits for higher time-step count must simulate more total time, circuits tend to grow in size with increasing simulation time-step \cite{wiebe11, smith2019simulating}. In fact, the "no-fast-forwarding theorem" states that simulating the dynamics of a quantum system under a generic Hamiltonian for time $T$ will require $\Omega(T)$ gates \cite{berry2007efficient, childs2009limitations}, implying circuit depths grow at least linearly with the number of time-steps.  On current NISQ hardware, there are limits to how large a circuit can get before qubit decoherence and gate-error rates reduce the fidelity of the simulation results, thus limiting the number of time-steps that can be simulated for a generic Hamiltonian.  It should be noted, however, that special classes of Hamiltonians (e.g., quadratic Hamiltonians) can be fast-forwarded, meaning that circuit depths need not grow significantly with simulation time \cite{atia2017fast}.  Recent work used a variational approach to fast-forward the dynamic simulation of several quadratic Hamiltonian models \cite{cirstoiu2020variational}.

\subsection{Magnetization}
One of the most straightforward observables to measure on current NISQ computers is the average magnetization of a spin system, as it only requires measuring the expectation value of a Pauli operator along the desired axis.  For example, to measure the average magnetization in the $x$-direction, the expectation value $\langle\sigma_x\rangle$ must be measured on each qubit and the resulting values averaged over all qubits.  For this reason, many dynamic simulations that have been executed on NISQ computers involve time evolution of spin systems under Hamiltonians that model magnetism in materials.  The last few years have seen numerous demonstrations of simulating the dynamic magnetization of the transverse field Ising model on NISQ computers including one simulating the non-equilibrium dynamics of quenches \cite{zhukov2018algorithmic}, one using out-of-equilibrium thermal states \cite{lamm2018simulation}, one using the TFIM to model a two-dimensional material \cite{bassman2020towards}, one using the Floquet formalism \cite{kyriienko2018floquet}, one using Jordan--Wigner, Fourier, and Bogoliubov transformations to diagonalize the Hamiltonian \cite{cervera2018exact}, one using the quantum Lanczos algorithm \cite{yeter-aydeniz2020Scattering}, and one using a hybrid classical-quantum method to utilize crosstalk between gates as analog fields \cite{babukhin_hybrid_2020}.  Ref. \cite{vovrosh2021efficient} showed how to use results from simulations of the dynamic local magnetization of the TFIM an additional longitudinal field to compute mesonic masses.  Dynamic simulations of other models derived from the Heisenberg model, including the XX and XXZ spin-chains, have also been carried out on NISQ computers \cite{smith2019simulating}.

\subsection{Dynamical Correlation Functions}
\begin{figure}[htpb]
    \includegraphics[clip=true,trim=80 650 100 100] {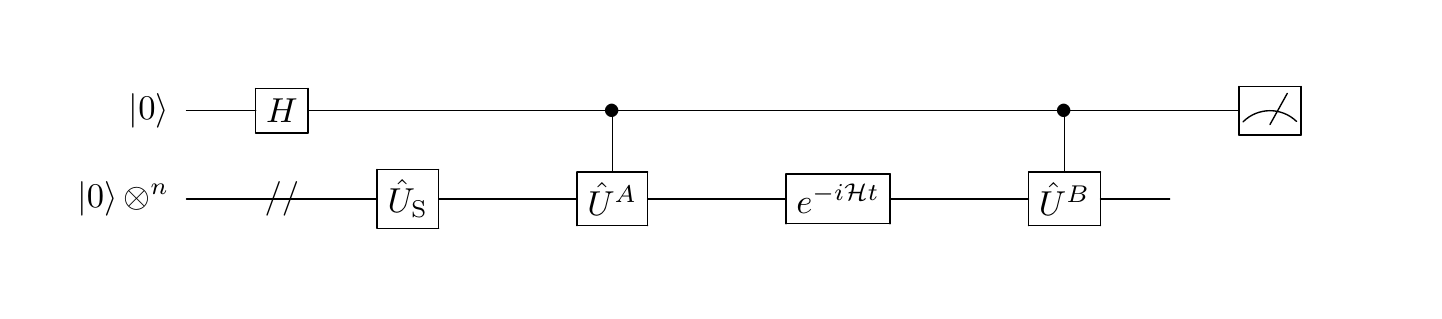}
    \caption{Generic circuit structure for measuring two-point correlation functions. The system qubits are prepared into a ground state via the operator $\hat{U}_S$.
    $\hat{U}^A$ and $\hat{U}^B$ encode the unitaries that represent the operators of the correlation function, which is measured in the
    coherences of the ancilla qubit.}
    \label{fig:corrfcn}
\end{figure}

Aside from single-time observables, it is natural to consider two-time correlation functions. These are typically associated with excitations of a material and correspond to observables such as neutron scattering and conductivity, or, for example, reveal the dynamics of spin waves. The current research has relied on a general formalism developed by Pedernales et al.~\cite{pedernales2014Efficient}, who describe how an extension of the Hadamard test can be used to compute general $n$-point correlation functions. The essential circuit is shown in Figure~\ref{fig:corrfcn}, and we briefly describe its operation for a correlation function of the form $\langle \hat{A}(t) \hat{B}(0)\rangle$.  After preparation of the ground state in some fashion, the $\hat{B}$ operator is applied, controlled on an ancilla in a superposition state. This ``splits'' the system into a ground state and an excited state, which are then subsequently time evolved to time $t$ under the system Hamiltonian. A final controlled application of the $\hat{A}$ operator produces the real and imaginary parts of the desired correlation function to be measurable in the coherent part of the ancilla qubit. In essence, this is a direct evaluation of the Lehmann representation of an operator.  This method has been used within the context of spin systems to compute the magnetic response of the Heisenberg molecule \cite{chiesa2019quantum} and chain at zero \cite{francis2020quantum} and finite temperatures \cite{sun2020quantum}.  The same approach is also used to measure Green's functions \cite{kreula2016few}, which is a critical ingredient for embedding methods (see Section~\ref{subsec:embedded}).

In the above, it is assumed that a correct ground state can be produced, and that the time evolution can be efficiently implemented to avoid Trotterization errors. However, correlation functions have an advantage in that the frequency content of the measurement is typically limited; in other words, the response of the system can only occur at the differences between the energy eigenvalues of the Hamiltonian. Moreover, the interest tends to be in low-frequency response. Thus, an application of a Fourier filter (either as a direct filter or by performing Fourier fits if the number of frequencies is known) enables the reduction of noise effects for finding frequencies. This is unfortunately not applicable for the signal amplitude, but there, sum rules or conservation laws can often be used.  Zero-time correlation functions have known values, which can be used to scale the measured response in some cases \cite{chiesa2019quantum,francis2020quantum,sun2020quantum}.

\subsection{Non-equilibrium Dynamics}
The ability to apply time evolution operators quite naturally enables the study of non-equilibrium dynamics of a system. This approach comes in a few flavors. First, the system may be prepared in the ground state of an initial Hamiltonian, and then time evolved under a final Hamiltonian, a process known as a quantum quench. Time-local measurements such as magnetization or densities are then used to characterize the state of the system as a function of time. This approach was recently demonstrated for the Markus model \cite{ollitrault2020non}, the Fermi--Hubbard model \cite{arute2020Observation}, and various subgroups of the Heisenberg model \cite{smith2019simulating,fauseweh2020digital}.  It has also been used to study various phenomena in the transverse field Ising mode including the dephasing of the model with long-range interactions \cite{kaplan2020many}, confinement and entanglement dynamics \cite{vovrosh2020Confinement, vovrosh2021efficient}, and dynamical phase transitions \cite{zhang2017observation}.

Another approach is to have explicit time-dependence in the system Hamiltonian. As discussed in Section~\ref{subsec:ham_evo}, since the Hamiltonian does not commute with itself at different times the time evolution operator (which is time-ordered) needs to be broken up into small time-steps $\Delta t$, with the approximating assumption that the Hamiltonian is constant over $\Delta t$. This approach has been applied to simulate the non-equilibrium dynamics of spin systems \cite{fauseweh2020digital, bassman2020towards}. Finally, the inherent noise in the quantum computer may be leveraged as a thermalizer as the system is propagated forward in time, which was applied to a study of small molecules \cite{rost2020noisy}.

Finally, non-equilibrium dynamics can be studied in open quantum systems, where the system and its environment are explicitly simulated.  A recent work simulated the dynamic population probabilities in a dissipative Hubbard Model, where separate qubit registers were used to represent the system and the environment, which was modeled as a spin-bath in thermal equilibrium \cite{tornow2020non}.

\subsection{Other}
Dynamic simulations have also been used to study more exotic phenomena.  For example, the separated dynamics of charge and spin in the Fermi--Hubbard model was recently observed \cite{arute2020Observation}.  Dynamic simulations were also utilized to study scattering in the transverse field Ising model \cite{gustafson2019quantum}.  Finally, the dynamics of the braiding of Majorana zero modes were simulated, which can give insights into improving topological quantum computers.

\section{Working Examples of Static and Dynamic Simulations}
\label{sec:examples}
In this section we provide two examples for how to map a materials simulation problem onto a quantum computer.  The first example shows how the problem of solving the Bardeen–Cooper–Schrieffer (BCS) gap equation \cite{bardeen1957theory} can be formulated as a static simulation on a quantum computer.  This is a prototypical problem of an interacting material system that contains a self-consistently determined parameter that may be solved variationally.  The second example demonstrates a how to set up a simulation of the non-equilibrium dynamics of a material on a quantum computer.  Such simulations can provide insights for fundamental questions about phase transitions, quantum critical points, equilibration, and thermalization in quantum materials.  Code for both working examples may be found in the form of Python notebooks in the Supplementary Material \cite{github_repo}.

\subsection{Static Simulation}
The BCS problem is a well-established model for the low-energy sector that is appropriate to both electron-phonon and electron-electron superconductivity. Solving this model for general systems is critical in determining the superconducting phase diagram in a host of materials, ranging from simple metals such as Pb to the high-T$_C$ cuprates and pnictides \cite{norman2008trend}.  More generally, however, it is representative of a model that has a static property that needs to be determined self-consistently; in this case it is the superconducting gap $\Delta$, but other extensions include magnetization and bipartite fields.  We will start with a brief overview of the problem to be solved, following the discussion in Capone et al.~\cite{Capone2018}. The gap equation arises from the so-called BCS Hamiltonian
\begin{eqnarray}
\mathcal{H}_{\mathrm{BCS}} = \sum_{\kk,\sigma} \epsilon_kk c^\dagger_{\kk,\sigma} c_{\kk,\sigma} - U \sum_{\kk,\pp}
c^\dagger_{\kk,\uparrow} c^\dagger_{-\kk,\downarrow}
c^\dagger_{-\pp,\downarrow}c^\dagger_{\pp,\uparrow} 
\end{eqnarray}
where $c^\dagger_{\kk,\sigma}$ ($c_{\kk,\sigma}$) creates (annihilates) a quasi-particle with momentum $\kk$ and spin $\sigma$. The quasi-particles have a non-interacting dispersion $\epsilon_\kk$ and experience an attractive interaction in the pairing channel with amplitude $U$. Here, we have neglected any momentum dependence in the interactions, which will lead to an $s$-wave solution.  In the mean field limit, the Hamiltonian becomes
\begin{eqnarray}
\mathcal{H}_{\mathrm{MF}} =  \sum_{\kk,\sigma} \epsilon_kk c^\dagger_{\kk,\sigma} c_{\kk,\sigma} 
- \sum_\kk \Delta c^\dagger_{\kk,\uparrow} c^\dagger_{-\kk,\downarrow} + \mathrm{H.C.}
\end{eqnarray}
with the mean field amplitude, or superconducting order parameter
\begin{eqnarray}
\Delta = \frac{U}{N_\kk} \sum_{\kk} \langle c_{-\kk,\downarrow} c_{\kk,\uparrow} \rangle =  \frac{U}{N_\kk}  \sum_{\kk} \langle c^\dagger_{\kk,\uparrow} c^\dagger_{-\kk,\downarrow} \rangle,
\end{eqnarray}
where $N_\kk$ is the number of unit cells.
The typical approach is to solve this self-consistent problem via a variational method or simple numerical self-consistency.

Here, we wish to map this problem onto qubits. For the mean-field solution of the attractive Hubbard model, this is made particularly
easy by the complete separation into momenta; the mean field is an independent sum of the individual contribution at each momentum.
As a result, the Hilbert space is decomposed into a product of momenta $\kk$ which each span a small Fock space $\mathcal{F}$:
\begin{eqnarray}
\mathcal{F} = \left\lbrace |0\rangle, |k\uparrow\rangle, |k\downarrow\rangle, |k\uparrow, k\downarrow\rangle \right\rbrace
\end{eqnarray}
This simplification enables the use of a particularly useful viewpoint by Anderson~\cite{anderson1958random},
known as the Anderson pseudospin representation, where combinations of fermionic bilinear operators are mapped onto operators
in $SU(2)$. The relevant ones here are
\begin{eqnarray}
S_{\kk}^x &= \frac{c^\dagger_{\kk,\uparrow} c^\dagger_{-\kk,\downarrow} + c_{-\kk,\downarrow} c_{\kk,\uparrow} }{2},\\
S_{\kk}^z &= \frac{1}{2} \sum_\sigma c^\dagger_{\kk,\sigma} c_{\kk,\sigma}.
\end{eqnarray}
Notice that the $S_\kk^z$ operators correspond to the occupation $n_\kk$, and $S_\kk^x$ to the contribution to the
mean field gap.

With these operators, the Hamiltonian for a particular momentum $\kk$ and the self-consistent equation for the gap $\Delta$ are
\begin{eqnarray}
\langle \mathcal{H} \rangle &= 2\sum_\kk \epsilon_\kk \langle S^z_\kk \rangle - \Delta \langle S^x_\kk \rangle,
\label{eq:pseudoBCS1}\\
\Delta &= \frac{U}{N_\kk} \sum_\kk \langle S^x_\kk \rangle.
\label{eq:pseudoBCS2}
\end{eqnarray}

\begin{figure}[h]
\centering
\hspace{0.2in}
\Qcircuit @C=1em @R=.7em 
{
\lstick{\ket{0}} & \gate{R_y(2\theta_\kk)} & \qw & \meter
}\\
\vspace{0.1in}
\includegraphics[width=0.49\textwidth]{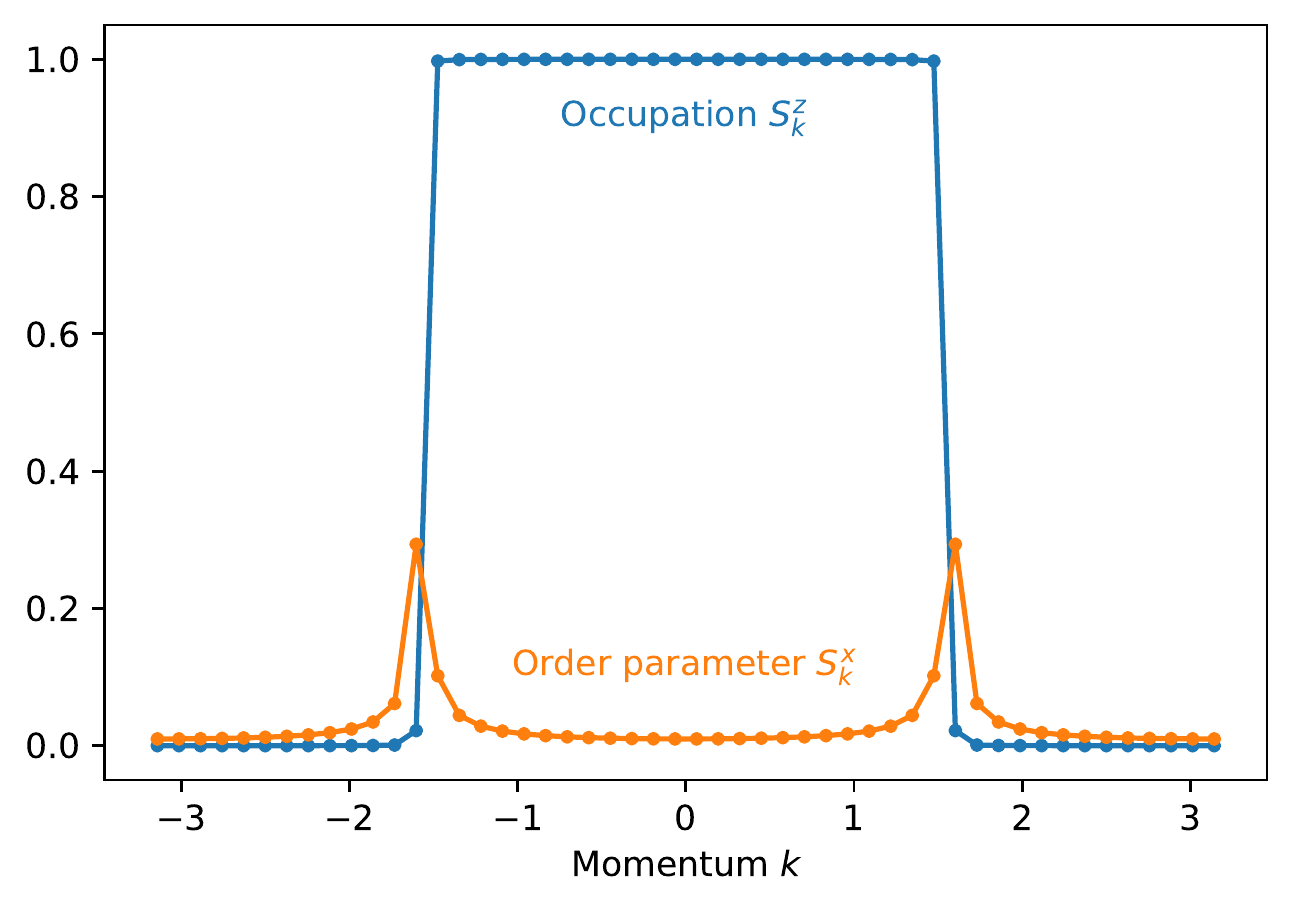}
\caption{Top: Quantum circuit for solving the BCS gap equation. Bottom: Result for a nearest neighbor hopping band structure with $U/t = 0.3$.}
\label{fig:bcs_circuit}
\end{figure}

We now have a simple optimization problem in hand; for each $\kk$ point there is an optimal (pseudo)spin direction in the $x-z$ plane,
as determined by the self-consistent equations (\ref{eq:pseudoBCS1}) and (\ref{eq:pseudoBCS2}).  Each momentum $\kk$ is mapped
to a single qubit, which is rotated to somewhere in the $x-z$ plane; the $x$ and $z$ projections correspond to the local contribution to the
gap, and to the occupation, respectively. The optimization parameter is thus the angle $\theta_\kk$ by which the qubit should be rotated.
As an initial guess, we may use the occupations:
a state with $|\kk|<k_F$ is occupied, and $\theta_\kk=0$; similarly,
a state with $|\kk|>k_F$ is empty, and $\theta_\kk=\pi/2$.
We may use the Hamiltonian as a simple cost function 
$\mathcal{C}\left\{\theta_\kk\right\} = \sum_{\kk} \left(2 \epsilon_\kk S^z_\kk  - \Delta S^x_\kk\right)$
that can be optimized using an appropriate optimizer. Here, the quantum circuit is straightforward (see Figure~\ref{fig:bcs_circuit}), as it is
just a simple rotation about the $y$-axis.
The complexity arises due to the self-consistency condition; $\Delta$ involves a sum over all $\kk$, and thus all momenta need to be evaluated
for each self-consistent step. Figure~\ref{fig:bcs_circuit} shows the results for a cosine band structure in 1D, with the attractive interaction $U/t=0.3$.

\subsection{Dynamic Simulation}
The most straightforward method to drive a quantum material out of equilibrium is through a quantum quench.  Quenches occur in materials with a sudden (non-adiabatic) change in their environment.  These can be simulated on a quantum computer by initializing the qubits into the ground state of an initial Hamiltonian $H_i$, and then simulating the time-evolution of the system under a final Hamiltonian $H_f$.  This abrupt change in the Hamiltonian models an analogous change in the environmental parameters, for example, an external field suddenly being turned on.  

In this example, we study the dynamics of a quantum quench of a one-dimensional (1D) antiferromagnetic (AF) Heisenberg model.  The Heisenberg model captures the behavior of a variety of quantum materials, including magnetic crystals \cite{hammar1999characterization, manousakis1991spin, lake2013multispinon}, low-dimensional magnets \cite{greven1994spin, woodward2002two}, and two-dimensional layered materials \cite{frey2018tuning}.  Quenches of this model may therefore provide insights into the non-equilibrium dynamics of numerous quantum materials.  

The Hamiltonian of interest is given by 
\begin{equation}
H(t) = J\sum_{i=1}\{\sigma_{i}^{x}\sigma_{i+1}^{x} + \sigma_{i}^{y}\sigma_{i+1}^{y} + g\sigma_{i}^{z}\sigma_{i+1}^{z} \}
\label{AF_Heisenberg}
\end{equation}
where the magnitude of the strength of the exchange couplings $J$ and $g$ can be tuned dynamically (requiring $J > 0$ and $g > 0$ makes this an AF model).  The qubits are initialized in the N\'eel state, given by $|\psi_0\rangle = |\uparrow \downarrow \uparrow ... \downarrow\rangle$, which is the ground state of the Hamiltonian in equation (\ref{AF_Heisenberg}) in the limit of $g \rightarrow \infty$. In this limit, and setting $J=1$, our initial Hamiltonian can be written $H_i(t<0) = C \sum\sigma_{i}^{z}\sigma_{i+1}^{z}$, where $C$ is an arbitrarily large constant, and our final Hamiltonian can be written as $H_f(t \geq 0) = \sum_i\{\sigma_{i}^{x}\sigma_{i+1}^{x} + \sigma_{i}^{y}\sigma_{i+1}^{y} + g\sigma_{i}^{z}\sigma_{i+1}^{z}\}$.  The observable of interest is the staggered magnetization (the square of which gives the AF order parameter) \cite{barmettler2010quantum}, which is defined as
\begin{equation}
m_s(t) = \frac{1}{N}\sum_i (-1)^i \langle\sigma_{i}^{z}(t)\rangle
\label{order_param}
\end{equation}
where $N$ is the number of spins in the system.  

To study the dynamics of the staggered magnetization using a quantum computer, a different quantum circuit must be created for each time-step $T$, which simulates the time-evolution of the system from time $t=0$ to time $\Delta_t T$, where $\Delta_t$ is the size of the time-step.  Each circuit must first initialize the qubits into the N\'eel state, which is straightforward as this is a simple product state.  On current NISQ computers, qubits all start in the "spin-up" orientation, so creating the N\'eel state only requires applying the $X$ gate to every other qubit to initialize this AF ground state.  

\begin{figure}
\centering
\hspace{0.5in}
\includegraphics[scale=0.65]{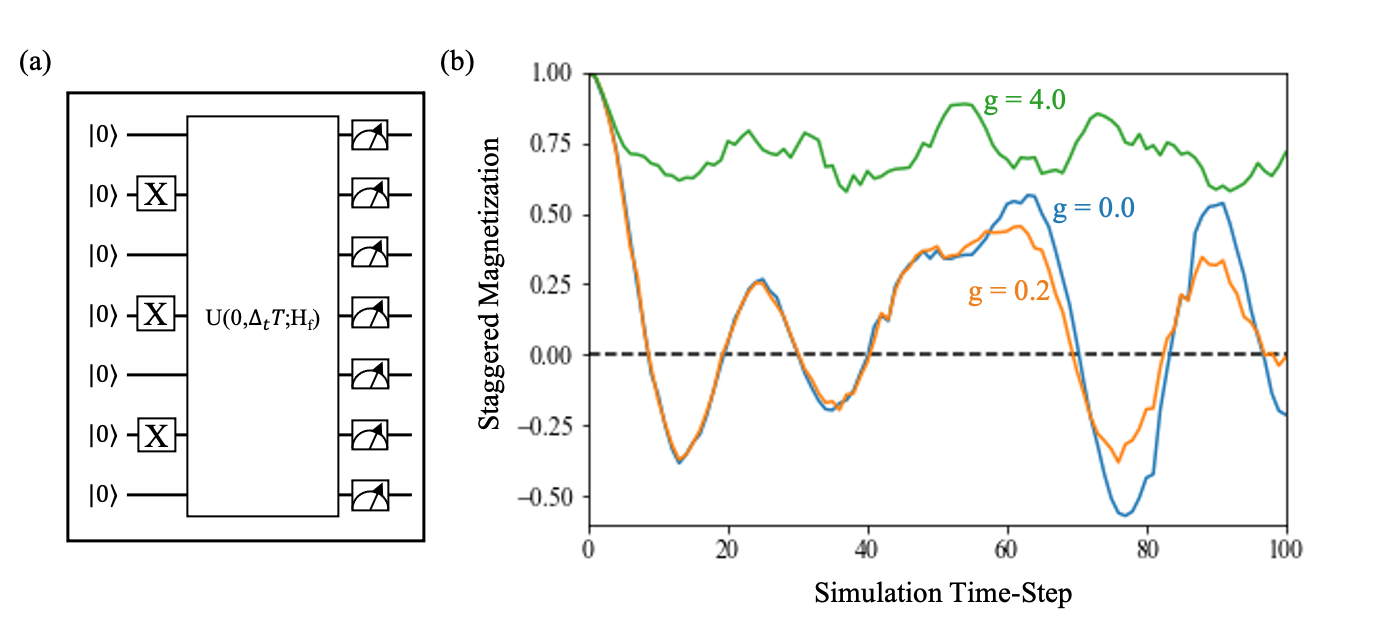}
\caption{Quantum circuit and simulation results for a quantum quench of the 1D AF Heisenberg model.  (a) A high-level quantum circuit diagram for a circuit simulating time-step $T$.  The $X$-gates on alternating qubits initialize the system into the N\'eel state.  The middle block labeled by the unitary operator $U(0,\Delta_tT;H_f)$ evolves the system from time $t=0$ to time $t=\Delta_tT$ under the Hamiltonian $H_f$, which is defined by the desired final value of $g$ in equation (\ref{AF_Heisenberg}).  Finally, measurement operators measure each qubit along the $z$-axis.  (b) Dynamics of the staggered magnetization after a quantum quench for three different values of $g$ in the final Hamiltonian.}
\label{fig:stag_mag}
\end{figure}

Next the time-evolution operator $U(0,\Delta_t T; H_f)$ must be converted into a set a gates to simulate evolution of the spins under the final Hamiltonian from time $t=0$ to time $T$.  This can be accomplished by using the Trotter decomposition to approximate $U$, which involves splitting the Hamiltonian into components that are each easily diagonalizable and breaking the total evolution time down into small time-steps.  In this case, $H_f$ can be broken down into three parts $H_f = H_x + H_y + H_z$ where $H_x = \sum_i\sigma_{i}^{x}\sigma_{i+1}^{x}$, $H_y = \sum_i\sigma_{i}^{y}\sigma_{i+1}^{y}$, $H_z = g\sum_i\sigma_{i}^{z}\sigma_{i+1}^{z}$, which gives the following approximation for the time-evolution operator
\begin{equation}
U(0,\Delta_tT) = \prod_{t=0}^T e^{-iH_x\Delta_t}e^{-iH_y\Delta_t}e^{-iH_z\Delta_t}
\label{time_evol}
\end{equation}
As each exponent can easily be converted into a set of one- and two-qubit gates, a quantum circuit implementing the operator in equation (\ref{time_evol}) can be created for each time-step by incrementally increasing the integer $T$.  By prefixing these circuits with the state initialization gates, and postfixing them with a measurement operator, a set of complete quantum circuits can be composed that will enable the dynamic simulation of the AF order parameter in the quench of a 1D Heisenberg model.  A high-level quantum circuit diagram for these circuits is depicted in Figure \ref{fig:stag_mag}a.
Upon running this set of circuits on a NISQ backend, minimal post-processing is required to compute the value of the staggered magnetization $m_s(t)$ from the values of $\langle\sigma_i^z\rangle$ that are returned for each qubit $i$ for each time-step.  Figure \ref{fig:stag_mag}b shows results for this dynamics simulation with a system consisting of seven spins with various values for $g$ in $H_f$.

\section{Summary and Outlook}
\label{sec:conclusion}
The holy grail for digital quantum computing is enabling scientific simulations of complex quantum materials that are intractable with classical computing resources, otherwise known as quantum advantage. At the time of writing of this review, we are not yet there, although trends in progress made over the last decade indicate that we getting closer.  The amassed literature shows that through the steady development of various methods and techniques, it has become possible to explore a plethora of static and dynamic properties of quantum materials on digital quantum computers (DQCs) for small and simplified models.  While all simulations of quantum systems performed on a quantum computer to date are still accessible to classical computers, these proof-of-concept simulations demonstrate that materials simulations on quantum computers are possible and illuminate the hurdles that must still be overcome.  

Moving to the realm of materials that are classically inaccessible will require advances in a number of different domains.  Quantum hardware can be scaled to larger qubit-counts with higher fidelities for qubits and quantum gates.  Algorithms can be improved to either tolerate more noise or reduce required quantum resources.  Software for programming quantum simulations can target higher layers of abstraction to make it easier and more efficient for scientists from various domains to contribute to progress.  New and improved encodings of material systems into qubits can assist in making the most of limited quantum resources.   Finally, much can still be learned from running proof-of-concept simulations of static and dynamic properties on both quantum simulators (with and without noise) as well as real quantum processors.  In this review, we have attempted to provide a snapshot of the current progress in each of these realms: hardware, software, algorithms, encodings, and successfully executed static and dynamic simulations of materials on quantum backends.  Clearly, the quest towards using DQCs to drive scientific discovery in materials science is a multi-disciplinary pursuit.  Scientists from physics, chemistry, materials science, and computer science can all make meaningful contributions.  As such, we aimed to make this review accessible to these diverse scientific domains and sought to provide a broad perspective, along with the tools and techniques, required to study quantum computation for materials simulations.  

While the latest DQCs are moving towards having the necessary number of qubits to encode materials problems beyond the abilities of classical resources, large scale simulations are still hampered by quantum noise presented in current quantum computers. Errors introduced by noise drastically limit the effective lifetime of a coherent quantum state on a quantum computer, and thus limit the number of quantum operations that can be performed reliably. Limiting the impact of or removing errors can be achieved through quantum error correction (QEC)~\cite{PhysRevA.54.1098, PhysRevLett.77.793}. However, the fault tolerance provided by QEC algorithms requires hundreds of thousands to millions of qubits, many orders of magnitude beyond the current quantum computing hardware capabilities. 

In the near-term, incremental progress in quantum hardware and theoretical advances, such as the development of more quantum resource efficient QEC schemes or Hamiltonian encoding approaches, will start to enable the materials resource community to explore ever more complex quantum simulations on quantum computing resources.  Over the longer term, more revolutionary advances are needed to either devise noisy quantum hardware on a scale large enough to support QEC, or gain a better understanding of and control over quantum noise in quantum hardware.  The latter field of research is itself just an investigation into the properties and behavior of quantum materials (which comprise the quantum hardware), and thus can be aided by DQCs.  Beyond speedups, quantum computers have the ability to simulate the complex Hamiltonians of qubits as they interact with their environment, as was envisioned by Feynman~\cite{feynman1982simulating}. Varying Hamiltonian and environmental parameters, which is for the most part straightforward to do in a simulation, but is much harder to do experimentally, can provide researchers with the essential insights needed to optimize quantum systems by reducing noise and increasing coherence and operational fidelities. In this sense, we see that enabling simulations of quantum materials on near-term DQCs directly aids in the design of new materials for future quantum computers (i.e., using quantum computers to design better quantum computers).

\section*{Acknowledgments}
\label{ack}
LB, MU, JC and WAdJ were supported by the U.S. Department of Energy (DOE) under Contract No.  DE-AC02-05CH11231,  through the Office of Advanced Scientific Computing  Research  Accelerated  Research  for  Quantum Computing  and  Quantum Algorithms Team Programs.” MM was partially supported by the “Embedding Quantum Computing into Many-body Frameworks for Strongly Correlated Molecular and Materials Systems” project, which is funded by the U.S. Department of Energy (DOE), Office of Science, Office of Basic Energy Sciences, the Division of Chemical Sciences, Geosciences, and Biosciences.  AFK was supported by the Department of Energy, Office of Basic Energy Sciences, Division of  Materials Sciences and Engineering under Grant No. DE-SC001946.

\newpage
\bibliographystyle{iopart-num}

\bibliography{qst}

\end{document}